\documentclass[11pt]{article}
\usepackage{jheppub}
\usepackage{graphicx}
\usepackage{amsmath, amssymb}
\usepackage{tikz}
\usepackage{MnSymbol}
\usepackage{hyperref}
\usepackage{color}
\usepackage[utf8]{inputenc}
\usetikzlibrary{arrows.meta,calc,decorations.markings,math,arrows.meta}

\def\vth{\vartheta}
\def\ovth{\overline{\vartheta}}

\def\re{\mathop{\hbox{\rm Re}}\nolimits}
\def\im{\mathop{\hbox{\rm Im}}\nolimits}

\newcommand{\ds}{\displaystyle}
\newcommand{\q}{{\mathsf q}}
\newcommand{\p}{{\mathsf p}}
\def\EXP{\textrm{{\large e}}}
\newcommand{\ii}{\mathsf{i}}

\newcommand{\olW}{\overline{W}}

\newcommand{\Sd}{S}

\newcommand{\sgn}{\varepsilon}
\newcommand{\nomea}{\tau_1}
\newcommand{\nomeb}{\tau_2}
\newcommand{\nomec}{\tau}
\newcommand{\spin}{\sigma}
\newcommand{\sspin}[1]{\mathbf{#1}}
\newcommand{\dspin}{a}

\newcommand{\bw}[3]{W\left(#1\,|\,#2,#3\right)}
\newcommand{\olbw}[3]{\olW\left(#1\,|\,#2,#3\right)}
\newcommand{\legf}[2]{\Phi(#1,#2)}
\newcommand{\legfb}[4]{\Phi(#1,#2\,;\,#3,#4)}
\newcommand{\rfac}[2]{R(#1,#2)}
\newcommand{\hatx}[3]{\hat{#1}}
\newcommand{\hatxb}[3]{\hat{#1}}
\newcommand{\lag}[3]{\mathcal{L}(#1\,|\,#2,#3)}
\newcommand{\olag}[3]{\overline{\mathcal{L}}(#1\,|\,#2,#3)}
\newcommand{\leg}[3]{\Psi(#1\,|\,#2,#3)}

\def\be#1\ee{\begin{align}#1\end{align}}

\newcommand{\bs}{\begin{split}}
\newcommand{\es}{\end{split}}

\title{Lens elliptic gamma function solution of the Yang-Baxter equation at roots of unity}

\author[a]{Andrew P. Kels}
\author[b]{and Masahito Yamazaki}

\affiliation[a]{Institute of Physics, University of Tokyo, Komaba, Tokyo 153-8902, Japan}
\affiliation[b]{Kavli IPMU (WPI), University of Tokyo, Kashiwa, Chiba 277-8583, Japan}

\emailAdd{andrew.p.kels(at)gmail.com}
\emailAdd{masahito.yamazaki(at)ipmu.jp}

\preprint{IPMU17-0127}

\abstract{
We study the root of unity limit of the lens elliptic gamma function solution of the star-triangle relation, for an integrable model with continuous and discrete spin variables.  This limit involves taking an elliptic nome to a primitive $rN$-th root of unity, where $r$ is an existing integer parameter of the lens elliptic gamma function, and $N$ is an additional integer parameter.  This is a singular limit of the star-triangle relation, and at subleading order of an asymptotic expansion, another star-triangle relation is obtained for a model with discrete spin variables in $\mathbb{Z}_{rN}$.  Some special choices of solutions of equation of motion are shown to result in well-known discrete spin solutions of the star-triangle relation.  The saddle point equations themselves are identified with three-leg forms of ``3D-consistent'' classical discrete integrable equations, known as $Q4$ and $Q3_{(\delta=0)}$.  We also comment on the implications for supersymmetric gauge theories, and in particular comment on a close parallel with the works of Nekrasov and Shatashvili.
} 

\date{\today}

\begin{document}

\maketitle

\section{Introduction}

The star-triangle relation (STR) is one of the most important forms of the Yang-Baxter equation, for integrability of two-dimensional models of statistical mechanics \cite{Baxter:1982zz}.  These are models where spins are located at the vertices of a lattice, or more generally vertices of a planar graph, and interactions take place between nearest-neighbour spins that are connected by edges.  Many important integrable models fall into this class, including the Ising model \cite{Baxter:1982zz}, and chiral Potts model \cite{AuYang:1987zc,Baxter:1987eq}, and also more recent continuous spin master type models \cite{Bazhanov:2010kz}, which contain the discrete spin models in special limits.\footnote{See e.g.\ \cite{Bazhanov:2016ajm} for a recent review of such continuous and discrete spin solutions of the STR.}

Recently, a new master type solution of the STR has been discovered \cite{Yamazaki:2013nra,Kels:2015bda}, with spins that have a combination of continuous and discrete components.  We call this the {\it lens-elliptic gamma function solution} of the STR, since the Boltzmann weights of the model are expressed in terms of products of the so-called lens elliptic gamma function \cite{Benini:2011nc,Razamat:2013opa,Kels:2015bda}.  The resulting integrable model is one of the most general known solutions of the Yang-Baxter equation, containing all previously known solutions of the STR that involve single component spins \cite{Bazhanov:2010kz}.  The corresponding integrable model is labelled by a positive integer $r$, and also depends on two complex valued elliptic nomes $\p$ and $\q$, that enter into the definition of the lens elliptic gamma function.  The case of $r=1$ is equivalent to the ``master solution'' of the STR \cite{Bazhanov:2010kz}, that has continuous spins taking values in $[0,\pi]$, while for general $r>1$ the model considered here has additional discrete spin components taking values in $\mathbb{Z}_r$.

The Boltzmann weights of the model were originally obtained in the context of the Gauge/YBE correspondence \cite{Yamazaki:2012cp,Terashima:2012cx,Yamazaki:2013nra}, where starting with a four-dimensional $\mathcal{N}=1$ quiver gauge theory, a corresponding solution of the Yang-Baxter equation is obtained by computing supersymmetric indices; Yang-Baxter equality is guaranteed by Yang-Baxter duality, stating that a certain pair of $\mathcal{N}=1$ quiver gauge theories flow to the same fixed point in the long-distance limit.  The case of the $S^1\times S^3/\mathbb{Z}_r$ index \cite{Benini:2011nc} (which we call the lens index) with $SU(N_c)$ gauge groups, corresponds to an integrable model with spins in $\mathbb{R}^{N_c-1}$ and $(\mathbb{Z}_r)^{N_c-1}$ \cite{Yamazaki:2013nra},\footnote{For the case of $r=1$, the $S^1\times S^3$ partition function \cite{Kinney:2005ej,Romelsberger:2005eg} generates the $r=1$ solution of \cite{Bazhanov:2010kz,Zabrodin:2010qm,Spiridonov:2010em}, see e.g.\ \cite{Yamazaki:2012cp,Terashima:2012cx}. See also \cite{Spiridonov:2010em,Xie:2012mr,Yagi:2015lha,Yamazaki:2015voa,Gahramanov:2015cva,Bazhanov:2016ajm,Maruyoshi:2016caf,rarified,GahramanovKels,Yamazaki:2016wnu,Yagi:2017hmj,Kels:2017toi,Jafarzade:2017fsc} for some more related works on the relation between quiver gauge theories and integrable models.} which for $N_c>2$ doesn't satisfy the STR, but rather satisfies the star-star relation \cite{Yamazaki:2013nra,Kels:2017toi}, the latter being related to another form of the Yang-Baxter equation \cite{Baxter:1997tn}.  Only in the special case of $N_c=2$ is a STR satisfied \cite{Kels:2015bda}, which corresponds to the lens elliptic gamma function solution of the STR mentioned above.

In this paper we study the limit of the lens elliptic gamma function solution of the STR, where one of the elliptic nomes goes to a root of unity.  This is a singular limit of the lens elliptic gamma function, where its asymptotics can be expressed in terms of products of Jacobi theta functions.  Accordingly this is also a singular limit of the Boltzmann weights, and in Section \ref{sec:limit} we utilise a saddle point method to evaluate the STR.  At subleading order $O(1)$ of an asymptotic expansion of the STR, another form of the STR is obtained for an integrable lattice model which only has discrete spin components.  Specifically, it is found that the continuous component of the spin, is replaced by a sum over the saddle points labelled by integers $\mathbb{Z}_N$, while the discrete component of the spin remains unchanged.  Thus this limit results in an STR for a lattice model that depends on two types of discrete spins, which take values in $\mathbb{Z}_r$, and $\mathbb{Z}_N$, respectively.  After evaluating the STR on a saddle point, it is found (after a suitable change of variables and use of modular transformations of Jacobi theta functions) that these two types of spins are effectively described in terms of a single spin, that takes values in $\mathbb{Z}_{rN}$.  Thus the final expression for the STR which is given in Section \ref{sec:discrete}, corresponds to an integrable $\mathbb{Z}_{rN}$-state lattice model.

The latter $\mathbb{Z}_{rN}$ model, in fact turns out to essentially be equivalent to the $\mathbb{Z}_{N}$ model that was obtained in the $r=1$ case \cite{Bazhanov:2010kz}, up to the change of $N\to rN$ in the latter.  This connection is rather unexpected, considering that the subleading order $O(1)$ asymptotics of the lens elliptic gamma function, given in Section \ref{sec:limit}, are quite different for the respective cases of $r=1$, and $r>1$.  Thus identically to the $r=1$ case \cite{Bazhanov:2010kz}, special solutions of the saddle point equation for the general $r>1$ case considered in Section \ref{sec:discrete}, will result in the Kashiwara-Miwa \cite{Kashiwara:1986tu}, Fateev-Zamolodchikov \cite{Fateev:1982wi}, and chiral Potts models \cite{AuYang:1987zc,Baxter:1987eq}.  For the latter case, it is shown in Section \ref{sec:cpmodel} that for the particular choice of variables, the chiral Potts curve is the unique curve that arises in order to satisfy the saddle point equation.  This provides new insight into the appearance of the chiral Potts curve in the rapidity parameterisation of the chiral Potts model.  The saddle point equations themselves, are found to be identical to the equations found for the $r=1$ case \cite{Bazhanov:2010kz}, and are identified as three-leg forms of the classical ``3D-consistent'' discrete lattice equations known as $Q4$, and $Q3_{(\delta=0)}$ \cite{ABS} (the latter arises in the trigonometric limit).  This is a new example (for $r>1$) of the recently observed correspondence \cite{Bazhanov:2007mh,Bazhanov:2010kz,Bazhanov:2016ajm,Kels:2017fyt} between quantum integrable models that satisfy the Yang-Baxter equation, and classical integrable equations that satisfy the 3D-consistency condition.  In Section \ref{sec:gauge} the implications of this work to supersymmetric gauge theories is considered, along with parallels with the works of Nekrasov and Shatashvili.  Finally, several aspects for future work are discussed in the Conclusion.

\section{Definitions}

\subsection{Lens elliptic gamma function}

Define a positive integer parameter $r$, taking values
\begin{align}
\label{rparam}
r=1,2,\ldots,
\end{align}
and define two complex elliptic nomes as
\begin{align}
\p=\EXP^{\pi\ii\nomea}\,,\qquad\q=\EXP^{\pi\ii\nomeb}\,,\qquad\im(\nomea)\,,\im(\nomeb)>0\,.
\label{nomedef}
\end{align}

The lens elliptic gamma function \cite{Benini:2011nc,Razamat:2013opa} depends on the values of the elliptic nomes \eqref{nomedef}, a complex variable $z$, and an integer variable $m\in\{0,1,\ldots,r-1\}$, and is defined as \cite{Kels:2015bda}
\begin{align}
\begin{split}
\legfb{z}{m}{\p}{\q}&=\exp\left(\frac{\ii m(r-m)}{6r}(6z+\pi(r-2m)(\nomea-\nomeb-1))\right) \\
&\quad\times\Phi_1\left(z+\left(\frac{r}{2}-m\right)\pi \nomea\,;\, \p\q, \p^r\right) \,
 \Phi_1\left(z-\left(\frac{r}{2}-m\right)\pi \nomeb\,;\, \p\q, \q^r\right) \,,
\end{split}
\label{Phi_m_def}
\end{align}
where $\Phi_1(z;\p,\q)$ is the regular elliptic gamma function\footnote{Another notation in the literature is
$
\Gamma(x\,;\,\p,\q)=
\prod_{j,k=0}^{\infty}
\tfrac{1-x^{-1}\p^{j+1} \q^{k+1}}
{1-x\p^{j} \q^{k}} \;.
$
The two notations are related by 
$\Phi_1(z\,;\,\p, \q)=\Gamma(e^{2 \ii z} \p \q\,;\, \p^2, \q^2)$.}  \cite{Ruijsenaars:1997:FOA,Felder}
\begin{align}
\Phi_1(z\,;\,\p,\q) 
&= \prod_{j,k=0}^{\infty} \frac{
  1-\EXP^{2\ii z} \p^{2j+1} \q^{2k+1}
}
{
  1-\EXP^{ - 2\ii z} \p^{2j+1} \q^{2k+1}
}
= \prod_{j=0}^{\infty} \frac{
  (\EXP^{2\ii z} \p^{2j+1} \q\,;\, \q^2)_{\infty}
}
{
 (\EXP^{-2\ii z} \p^{2j+1} \q\,;\, \q^2)_{\infty}
}\,.
\label{Phi_1_def}
\end{align}
For $r=1$ the expression \eqref{Phi_m_def} is equivalent to \eqref{Phi_1_def}.

Due to the special choice of normalisation of \eqref{Phi_m_def},\footnote{Interestingly the normalisation of \eqref{Phi_m_def} can be expressed entirely in terms of multiple Bernoulli polynomials \cite{GahramanovKels}, but this property will not be used for this paper.} the function satisfies the important properties
\begin{align}
\legfb{z}{m}{\p}{\q}\,\legfb{-z}{-m}{\p}{\q}=1\,,\qquad\legfb{z}{m+kr}{\p}{\q}=\legfb{z}{m}{\p}{\q}\,,\quad k\in\mathbb{Z}\,,
\label{Phi_properties}
\end{align}
and
\begin{align}
\legfb{z+2\pi r}{m}{\p}{\q}=\legfb{z}{m}{\p}{\q}\,.
\end{align}

For brevity we often drop the elliptic nomes in the argument of $\legfb{z}{m}{\p}{\q}$, and write it as
\begin{align}
\legf{z}{m} :=\legfb{z}{m}{\p}{\q}\,.
\end{align}


\subsection{Boltzmann weights}

In the lattice model of statistical mechanics considered here \cite{Kels:2015bda}, we have a spin $\spin_i=(x_i, m_i)$ at each vertex $i$, with components that take values $x_i \in [0,\pi]$ and $m_i \in\{0,1,\ldots,r-1\}$.  The integration measure for this spin is defined as
\begin{align}
\sum_{\spin_0}:=\sum_{m_0=0}^{r-1}\,\int^\pi_0dx_0\,.
\label{STRmeasure}
\end{align}

The explicit expression for the edge Boltzmann weights, given in terms of the lens elliptic gamma function \eqref{Phi_m_def}, are
\begin{align}
\begin{split}
\bw{\alpha}{\spin_i}{\spin_j}&=
\frac{
\legf{x_i-x_j+\ii\alpha}{m_i-m_j}\, \legf{x_i+x_j+\ii \alpha}{m_i+m_j}
}{
\legf{x_i-x_j-\ii\alpha}{m_i-m_j}\, \legf{x_i+x_j-\ii \alpha}{m_i+m_j}
}
\,, \\[0.1cm]
\olbw{\alpha}{\spin_i}{\spin_j}&=\bw{\eta-\alpha}{\spin_i}{\spin_j}\,,
\end{split}
\label{bwdef_W_lens}
\end{align}
where $\eta$ is the crossing parameter, defined as
\begin{align}
\eta=-\frac{\pi\ii}{2}(\nomea+\nomeb)\,.
\label{etadef}
\end{align}
The Boltzmann weights \eqref{bwdef_W_lens} satisfy the symmetries
\begin{align}
\label{lensBSsym}
\bw{\alpha}{\spin_i}{\spin_j}=\bw{\alpha}{\spin_j}{\spin_i}\,,\quad \bw{\alpha}{\spin_i}{\spin_j}\,\bw{-\alpha}{\spin_i}{\spin_j}=1\,,
\end{align}
and are invariant under the shifts
\begin{align}
\label{shiftinvar}
x_i\rightarrow x_i+\pi\,,\quad x_j\rightarrow x_j+\pi\,,\quad m_i\rightarrow m_i+r\,,\quad m_j\rightarrow m_j+r\,,
\end{align}
as well as the combined shift
\begin{align}
\label{shiftinvar2}
(x_i,x_j,m_i,m_j)\to(-x_i,-x_j,-m_i,-m_j)\,.
\end{align}

The vertex Boltzmann weight is defined as
\begin{align}
\begin{split}
S(\spin_i)&=\frac{\EXP^{\frac{\eta r}{2}-\frac{8\eta m_i^2}{r}}}{\pi}\,
\vartheta_1(2(x_i-\pi m_i\nomea)\,|\,r\nomea)\,\vartheta_1(2(x_i+\pi m_i\nomeb)\,|\,r\nomeb) \\[-0.1cm]
&=\frac{G(\nomea r)\,G(\nomeb r)}{\legf{2x_i+\ii\eta}{2m_i}\,\legf{-2x_i+\ii\eta}{-2m_i}} \,,
\end{split}
\label{bwdef_S}
\end{align}
where $\vth_1(z\,|\,\tau)$ is the Jacobi theta function, defined in \eqref{theta_def}, and $G(\nomea)$ is a $q$-Pochhammer symbol, defined in \eqref{qpochdef}.  The vertex Boltzmann weight \eqref{bwdef_S} is a function of the single spin $\spin_i$ only, and does not depend on the value of a spectral variable $\alpha$.  It is also invariant under the shifts \eqref{shiftinvar}, \eqref{shiftinvar2}, for the components $x_i$ and $m_i$.

Finally the spin-independent normalisation factor $\rfac{\alpha_1}{\alpha_3}$ is defined as
\begin{align}
\rfac{\alpha_1}{\alpha_3}&=\frac{\legf{\ii(\eta-2\alpha_1)}{0}\,\legf{\ii(\eta-2\alpha_3)}{0}}{\legf{\ii(\eta-2(\alpha_1+\alpha_3))}{0}}\,.
\label{bwdef_R}
\end{align}
Note that the normalization factor $\rfac{\alpha_1}{\alpha_3}$, also has a special factorisation as
\begin{align}
\rfac{\alpha_1}{\alpha_3}=\frac{\kappa(\eta-\alpha_1)}{\kappa(\alpha_1)}
\frac{\kappa(\eta-\alpha_3)}{\kappa(\alpha_3)}
\frac{\kappa(\alpha_1+\alpha_3)}{\kappa(\eta-\alpha_1-\alpha_3)}
\,,
\label{rfactorisation}
\end{align}
where $\kappa(\alpha)$ is given for $0\leq\alpha<\eta$ by
\begin{align}
\kappa(\alpha):=&\exp \left\{
\sum_{n\ne 0} \frac{e^{\alpha n} ((\p \q)^{rn}-(\p \q)^{-rn}) }
{n((\p \q)^{2n}-(\p \q)^{-2n})(\p^{rn}-\p^{-rn})(\q^{rn}-\q^{-rn})}
\right\}
\,.
\label{kapnorm}
\end{align}
This function satisfies the pair of functional equations
\begin{align}
\frac{\kappa(\eta-\alpha)}{\kappa(\alpha)}=\Phi_{0}(\ii(\eta-2\alpha))\,,\qquad
\kappa(\alpha) \kappa(-\alpha)=1 \,.
\end{align}

The functions \eqref{bwdef_W_lens}, \eqref{bwdef_S}, \eqref{bwdef_R}, also depend implicitly on the elliptic nomes \eqref{nomedef}.

\subsection{Star-triangle relation}

The functions \eqref{bwdef_W_lens}, \eqref{bwdef_S}, \eqref{bwdef_R} define a solution of the following star-triangle relation
\begin{align}
\begin{split}
&\sum_{\spin_0}\,\,S(\spin_0)\,\olbw{\alpha_1}{\spin_0}{\spin_1}\,\bw{\alpha_1+\alpha_3}{\spin_0}{\spin_2}\,\olbw{\alpha_3}{\spin_0}{\spin_3} \\[-0.1cm]
&\qquad=\rfac{\alpha_1}{\alpha_3}\, \bw{\alpha_1}{\spin_2}{\spin_3}\,\olbw{\alpha_1+\alpha_3}{\spin_1}{\spin_3}\,\bw{\alpha_3}{\spin_1}{\spin_2}\,,
\end{split}
\label{STReq}
\end{align}
with integration measure given by \eqref{STRmeasure}.

In \eqref{STReq} there are five explicit independent variables $\spin_1,\spin_2,\spin_3$, and $\alpha_1,\alpha_3$; in addition the Boltzmann weights may depend implicitly on some additional parameters, such as a system temperature or elliptic nome, as is the case here.  For lattice models of statistical mechanics, the Boltzmann weights $\bw{\alpha}{\spin_i}{\spin_j}$, $\olbw{\alpha}{\spin_i}{\spin_j}$ are edge weights that characterise interactions between two spins $\spin_i, \spin_j$ connected by an edge of the lattice, $S(\spin_i)$ is a self-interaction weight associated to vertices of the lattice.  The sum in \eqref{STReq} is taken over the set of values assigned to a spin $\spin$.  In this case the sum is defined in \eqref{STRmeasure}, while in other cases the set is a discrete subset $\mathbb{Z}_N$ of integer spin states, or a continuous subset of values on the real line $[a,b]$, where in the latter case the measure is $\sum_{\spin_0}\rightarrow\int^b_ad\spin_0$.  Equation \eqref{STReq} is a typical form of the star-triangle relation for the majority of lattice models of statistical mechanics, with the notable exception of the chiral Potts model, which is presented in Section \ref{sec:cpmodel}.

As mentioned in the introduction, \eqref{STReq} also has interesting applications outside of integrable models of statistical mechanics.  For example, \eqref{STReq} may be directly interpreted as the Seiberg duality \cite{Seiberg:1994pq} of pairs of $\mathcal{N}=1$ $S^1\times S^3/\mathbb{Z}_r$ indices \cite{Benini:2011nc} of supersymmetric gauge theory, while in the area of elliptic hypergeometric functions, \eqref{STReq} is an identity that extends the elliptic beta integral \cite{SpiridonovEBF} to the case of both complex and integer variables, and appears as a simplest case of a transformation formula between elliptic hypergeometric ``sum/integrals'' on the $A_n$ and $BC_n$ root systems \cite{Kels:2017toi}.

\section{Root-of-unity limit}\label{sec:limit}

In this section, the following root of unity limit will be considered\footnote{Either of the elliptic nomes $\p$, or $\q$, may be taken to a root of unity.  The lens elliptic gamma function changes by a factor $\EXP^{\pi\ii m(r-m)(r-2m)/(3r)}$ under the simultaneous transformation $\q\leftrightarrow\p$, and $m\leftrightarrow r-m$ (while the STR \eqref{STReq} is invariant under this transformation), and so the difference between the two cases is superficial.}
\begin{align}
\p=\EXP^{\pi\ii\nomec}\,,\quad\q=\EXP^{-\frac{\hbar}{r}}\zeta \,,\qquad\hbar\rightarrow0^+\,,
\label{root_of_unity_limit}
\end{align}
with $\zeta$ being a primitive $2rN$-th root of unity ($\zeta=e^{\frac{2\pi \ii}{2rN}}$).  In \eqref{root_of_unity_limit}, we have set $\nomea\rightarrow\nomec$, since there is now only one $\nomec$ parameter to consider.

For the purposes here, it is convenient to introduce the following shifted variables
\begin{align}
\hatx{z}{z}{m}=z+\frac{\pi m}{rN}\,,\qquad\hat{\nomec}=N\nomec+\frac{1}{r}\,.
\label{varshiftdef}
\end{align}
The key equations for evaluating the star-triangle relation in the limit \eqref{root_of_unity_limit}, are the root of unity asymptotics of the lens elliptic gamma function, which may be expressed as
\begin{align}
\log\legf{z}{m}=\hbar^{-1}\Phi^{(-1)}(z,m)+\log\Phi^{(0)}(z,m) +O(\hbar)\,,
\label{key_equation}
\end{align}
where
\begin{align}
\Phi^{(-1)}(z,m)=\frac{\ii}{N}\int^{\hatx{z}{z}{m}}_0du\log\ovth_4\left(Nu\,|\,\hat{\nomec}\right)\,,
\label{lead}
\end{align}
and
\begin{align}
\begin{split}
\Phi^{(0)}(z,m)
&=\exp\left(\frac{\ii m(r-m)}{6r^2N}(6rNz+\pi(r-2m)(rN(\nomec-1)-1))\right) \\[0.0cm]
&\times\prod_{k_1=0}^{N-1}\,\prod_{k_2=0}^{r-1}\ovth_4\left(\hatx{z}{z}{m}+\frac{\pi}{N}(k_1+1)-\frac{\pi\hat{\nomec}}{2N}(2k_2+1)+\frac{\pi r\nomec}{2}\,\Big|\,r\nomec\right)^{\frac{rN-2r(k_1+1)+2k_2-2m+1}{2rN}} \\[0.1cm]
&\times\prod_{k=0}^{|m| -1}\,\ovth_4\left(\hatx{z}{z}{m}-\frac{\sgn(m)\pi\hat{\nomec}}{2N}(2k+1)+\frac{\pi r\nomec}{2}\,\Big|\,r\nomec\right)^{\sgn(m)},
\end{split}
\label{slead}
\end{align}
where $\ovth_4(z\,|\,\tau)$ is one of the Jacobi theta functions defined in \eqref{theta_def}, and the $\sgn$ function here is defined as
\begin{align}
\sgn(m)=\left\{\begin{array}{ll}
1\,,&\quad m\geq0\,, \\
-1\,,&\quad m<0\,. \end{array}\right.
\end{align}

The leading order asymptotics \eqref{lead}, come entirely from the singular factor $\Phi_1(z-(r/2-m)\pi\nomeb\,;\,\p\q,\q^r)$ in the definition of the lens elliptic gamma function \eqref{Phi_m_def}, while the other factors in \eqref{Phi_m_def} are $O(1)$.  Particularly, \eqref{slead} is equivalent to the product of the non-singular factor $\Phi_1(z+(r/2-m)\pi\nomea\,;\,\p\q,\p^r)$ in \eqref{Phi_m_def}, and the $O(1)$ contribution of the singular factor $\Phi_1(z-(r/2-m)\pi\nomeb\,;\,\p\q,\q^r)$ (and the $O(1)$ contribution of the exponential factor).  The final expression \eqref{slead} is obtained after rearranging the resulting infinite products into the form of Jacobi theta functions appearing in \eqref{theta_def}, while any remaining infinite products (not in the form of \eqref{theta_def}) end up cancelling out.

Note that at leading order, there is only dependence on $z$ and $m$ through the combination $\hatx{z}{z}{m}$, while at subleading order there is also a dependence on $z$ and $m$ that appears outside of this combination.  This means that $\hatx{z}{z}{m}$, may be taken as a new complex variable, with effectively no dependence on discrete spins at leading order.  Also the leading asymptotic term \eqref{lead} is invariant under a shift of $\hatx{z}{z}{m}\to\hatx{z}{z}{m}+\frac{\pi n}{N}$, where $n$ is an integer, while the subleading asymptotic term \eqref{slead} changes non-trivially under the same shift.  The $n\pi/N$ periodicity of the leading asymptotics will need to be taken into account when evaluating the star-triangle relation in the following section.

\subsection{Expansion of the star-triangle relation}

Since the leading asymptotics of the lens elliptic gamma function \eqref{lead} are $O(\hbar^{-1})$, the expansion of the star-triangle relation \eqref{STReq} will have the form
\begin{align}
\begin{split}
\sum_{m_0=0}^{r-1}\int^{2\pi}_0\!\frac{dx_0}{\sqrt{\hbar}}\,\,  &\exp\left(\frac{1}{\hbar} A_{\medstar}(\hatx{x}{x_0}{m_0}_0\,;\,\hatx{x}{x_i}{m_i}_i\,;\,\alpha_1,\alpha_3)+B_{\medstar}(x_0,m_0\,;\,x_i,m_i\,;\,\alpha_1,\alpha_3)\right) \\[0.1cm]
=&\exp\left(\frac{1}{\hbar} A_{\triangle}(\hatx{x}{x_i}{m_i}_i\,;\,\alpha_1,\alpha_3)+B_{\triangle}(x_i\,;\,m_i\,;\,\alpha_1,\alpha_3)\right) \left(1+O(\hbar) \right),
\end{split}
\label{STRlim}
\end{align}
where $A_{\medstar}(\hatx{x}{x_0}{m_0}_0\,;\,\hatx{x}{x_i}{m_i}_i\,;\,\alpha_1,\alpha_3)$, $A_{\triangle}(\hatx{x}{x_i}{m_i}_i\,;\,\alpha_1,\alpha_3)$, $B_{\medstar}(x_0,m_0\,;\,x_i,m_i\,;\,\alpha_1,\alpha_3)$, $B_{\triangle}(x_i\,;\,m_i\,;\,\alpha_1,\alpha_3)$, are independent of $\hbar$.  The factor of $1/\sqrt{\hbar}$ in \eqref{STRlim} comes from the calculation of the asymptotics of the factor $S(x)$ in the star-triangle relation \eqref{STReq} (this is an $O(\log \hbar)$ contribution inside the exponential).  Since they are constructed from \eqref{lead}, the leading asymptotic terms $A_{\medstar}(\hatx{x}{x_0}{m_0}_0\,;\,\hatx{x}{x_i}{m_i}_i\,;\,\alpha_1,\alpha_3)$, and $A_{\triangle}(\hatx{x}{x_i}{m_i}_i\,;\,\alpha_1,\alpha_3)$, depend on $x_i$, and $m_i$, only through the combination $\hatx{x}{x_i}{m_i}_i$, while the next order asymptotic terms $B_{\medstar}(x_0,m_0\,;\,x_i,m_i\,;\,\alpha_1,\alpha_3)$, and $B_{\triangle}(x_i\,;\,m_i\,;\,\alpha_1,\alpha_3)$, also depend on $x_i$ and $m_i$ outside of this combination, as in \eqref{slead}.

In the limit $\hbar\to 0$, the left hand side of \eqref{STRlim} can be evaluated with a saddle point method. The saddle point $\hatxb{x}{x_*}{m_*}_*$, is given by the solution to the equation
\begin{align}
\left.\frac{\partial A_{\medstar}(\hatx{x}{x_0}{m_0}_0\,;\,\hatx{x}{x_i}{m_i}_i\,;\,\alpha_1,\alpha_3)}{\partial \hatxb{x}{x_0}{m_0}_0} \right|_{\hatxb{x}{x_0}{m_0}_0=\hatxb{x}{x_*}{m_*}_*}=0\,.
\label{saddle_eq}
\end{align}
Due to the dependence only on the shifted variables $\hatx{x}{x_i}{m_i}_i$, the left hand side of \eqref{STRlim} is a sum of $r$ integrals, with $r$ different saddle points for $m_0=0,\ldots,r-1$, each satisfying \eqref{saddle_eq}.

The equation \eqref{saddle_eq} is evaluated for fixed choices of the variables $\hatxb{x}{x_i}{m_i}_i$, $\alpha_i$, and thus the saddle point $x_*$ in general depends on the values of $\hatxb{x}{x_i}{m_i}_i$, and $\alpha_i$.  It is also assumed here that there exists a solution to \eqref{saddle_eq}.  If there are multiple saddle points that give different values of $A_{\medstar}(\hatx{x}{x_0}{m_0}_0\,;\,\hatx{x}{x_i}{m_i}_i\,;\,\alpha_1,\alpha_3)$, then typically it is only needed to consider the saddle point which gives the largest absolute value of $A_{\medstar}(\hatx{x}{x_0}{m_0}_0\,;\,\hatx{x}{x_i}{m_i}_i\,;\,\alpha_1,\alpha_3)$, since the contributions from any other saddle points will become relatively smaller exponentially as $\hbar\rightarrow0$.  The situation here also has a subtlety due to the form of the leading asymptotics \eqref{lead}, where the saddle point equation \eqref{saddle_eq} is invariant under a shift $\hatxb{x}{x_0}{m_0}_0\to \hatxb{x}{x_0}{m_0}_0+\pi/N$, meaning that there are $N$ saddle points:
\begin{align}
\hatxb{x}{x_*}{m_0}_{*}=\zeta+\frac{\pi n}{N}\;, \qquad 0\leq\textrm{Re}\, \zeta < \frac{\pi}{N}\;, \qquad n=0, 1, \ldots, N-1 \,,
\label{saddle_m}
\end{align}
in the interval $[0,\pi]$, which give the same maximum value of $A_{\medstar}(\hatx{x}{x_*}{m_*}_*\,;\,\hatx{x}{x_i}{m_i}_i\,;\,\alpha_1,\alpha_3)$.  In this case the contribution of each saddle point needs to be considered, and the difference in the contributions only will manifest through the asymptotics at subleading order \eqref{slead}, where there is no longer an invariance under the shift $\hatxb{x}{x_0}{m_0}_0\to \hatxb{x}{x_0}{m_0}_0+\pi/N$.

Once the saddle points have been determined (some explicit solutions will be given in Section \ref{sec:discrete}), the left hand side of \eqref{STRlim} may be evaluated through a standard Gaussian integration as
\begin{align}
\frac{\exp\left(\frac{1}{\hbar}A_{\medstar}(\hatx{x}{x_*}{m_0}_*\,;\,\hatx{x}{x_i}{m_i}_i\,;\,\alpha_1,\alpha_3)+B_{\medstar}(x_*,m_0\,;\,x_i,m_i\,;\,\alpha_1,\alpha_3)\right)}{\left( \frac{1}{2\pi}\frac{\partial^2 A_{\medstar}(\hatx{x}{x_0}{m_0}_0\,;\,\hatx{x}{x_i}{m_i}_i\,;\,\alpha_1,\alpha_3)}{\partial \hatxb{x}{x_0}{m_0}_0^2}\Big|_{\hatxb{x}{x_0}{m_0}_0=\hatxb{x}{x_*}{m_*}_*} \right)^{\frac{1}{2}}}  \left(1+O(\hbar) \right)
\,.
\end{align}

Since the star-triangle relation \eqref{STReq} is an equality, the asymptotic expansion of both sides should be consistent at each order in $\hbar$.  This leads to new star-triangle type identities, that must hold at each order in $\hbar$.  In the case of \eqref{STRlim}, the following classical star-triangle relation is found at leading order $O(\hbar^{-1})$
\begin{align}
A_{\medstar}(\hatx{x}{x_*}{m_0}_*\,;\,\hatx{x}{x_i}{m_i}_i\,;\,\alpha_1,\alpha_3)=A_{\triangle}(\hatx{x}{x_i}{m_i}_i\,;\,\alpha_1,\alpha_3)\,,
\label{cSTR}
\end{align}
while the subleading order $O(\hbar^0)$ gives the star-triangle relation
\begin{align}
\begin{split}
\sum_{m_0=0}^{N-1}\left\{
B_{\medstar}\left(x_*+\frac{2\pi m}{N},m_0\,;\,x_i,m_i\,;\,\alpha_1,\alpha_3\right)  +\log\left( \frac{1}{2\pi}\frac{\partial^2 A_{\medstar}(\hatx{x}{x_0}{m_0}_0\,;\,\hatx{x}{x_i}{m_i}_i\,;\,\alpha_1,\alpha_3)}{\partial \hatxb{x}{x_0}{m_0}_0^2} \right)_{\hatxb{x}{x_0}{m_0}_0=\hatxb{x}{x_*}{m_*}_*+\frac{2\pi m}{N}}^{-\frac{1}{2}}\right\} \\
=B_{\Delta}(x_i\,;\,m_i\,;\,\alpha_1,\alpha_3) \,.
\label{dSTR}
\end{split}
\end{align}

In \eqref{dSTR}, there is a sum that is required to count the contributions from the $N$ saddle points\footnote{In general, to determine the relative weights for different saddle points we need to decompose the integration contour (which in this case is a real line) into an integer linear combination of Lefschetz thimbles corresponding to different saddle points, and the relative weights for the saddle points are determined by the integer coefficients (see e.g.\ \cite{Pham} and more recently \cite{Witten:2010cx}).
 In our problem it is natural to assume the $\mathbb{Z}_N$ ``replica symmetry'' exchanging different saddle points to be unbroken, and we will assume this throughout. This ensures that we should sum over different saddle points with the same weights.} coming from \eqref{saddle_m}.  Explicit forms of the equations \eqref{cSTR} and \eqref{dSTR} will be given in Section \ref{sec:discrete}.  A main result of this paper is showing that \eqref{dSTR} has the standard form of the star-triangle relation \eqref{STReq}, and special solutions of \eqref{saddle_eq} correspond to well-known integrable models of statistical mechanics.

\subsection{Change of variables}

Since the saddle point equations are invariant under the shift $\hatxb{x}{x_i}{m_i}_i\rightarrow \hatxb{x}{x_i}{m_i}_i+\pi k/N$, for integers $k$, in the following the spin will be redefined as
\begin{align}
\hatxb{x}{x_i}{m_i}_i=\zeta_i+\frac{\pi n_i}{N}\,,\qquad 0\leq\textrm{Re}\zeta<\frac{\pi}{N}\,,\qquad n_i=0,\ldots,N-1\,.
\label{ndef}
\end{align}
Also observe that in the limit \eqref{root_of_unity_limit}, the crossing parameter \eqref{etadef} is
\begin{align}
\eta=-\frac{\pi\ii}{2}\left(\nomec+\frac{1}{rN}\right)\,,\qquad \hbar\rightarrow0\,.
\end{align}
It turns out that it is more suitable here to analyse the equations \eqref{STRlim}, \eqref{saddle_eq}, in a set of variables where the (renormalised) crossing parameter is on the real line. For example, in the limit \eqref{root_of_unity_limit}, this may be done with the following change of variable
\begin{align}
\eta':= \frac{\ii rN\eta}{1+rN \nomec}= \frac{\pi}{2} \,.
\label{COVeta}
\end{align}
This then implies the following additional changes of variables in \eqref{STRlim}
\begin{align}
\phi_i=\left(\zeta_i+\frac{r\pi\nomec}{4}\right)\frac{rN}{1+rN\nomec}\,,\qquad\theta_i=\frac{\ii rN\alpha_i}{1+rN\nomec}\,,\qquad\nomec'=\frac{rN\nomec}{1+rN\nomec}\,.
\label{COV}
\end{align} 
Implementing the above change of variables, for the asymptotics given in \eqref{lead}, and \eqref{slead}, requires repeated use of the modular transformation identities for the Jacobi theta functions listed in Appendix \ref{app.Jacobi}.

There are several advantages of working in the variables $\phi_i$, $\theta_i$, $\tau'$, instead of the variables $x_i$, $\alpha_i$, $\tau$.  For example, following the change of variables \eqref{COV}, the Boltzmann weights of the star-triangle relation \eqref{dSTR} will be seen to have the usual ``additive'' form, and a physical regime for the corresponding model may be found by typically restricting the new variables $\phi_i$, $\theta_i$, to take values in a certain subset of the real line.  Furthermore the constant value of the crossing parameter \eqref{COVeta} is straightforwardly identified with the crossing parameter usually seen for integrable lattice models, which makes the comparison to known integrable models simpler.  Also with the new variables \eqref{COV}, some explicit solutions of the equation of motion \eqref{saddle_m} are found for special values of the $\phi_i$, and $\theta_i$, whereas it is typically non-trivial to find solutions of \eqref{saddle_m} for general values of the $\hatxb{x}{x_i}{m_i}_i$, $\alpha_i$.

\subsection{\texorpdfstring{Leading order ($O(\hbar^{-1})$) expansion}{Leading order (O(hbar**(-1)) expansion:}}

In terms of the variables $\phi_i$, $\theta_i$, defined in \eqref{COV}, the leading asymptotics of the star-triangle relation \eqref{STRlim} may be written as
\begin{align}
A_{\medstar}(\phi_0\,;\,\phi_1,\phi_2,\phi_3\,;\,\theta_1,\theta_3)&=\olag{\theta_1}{\phi_0}{\phi_1}+\lag{\theta_1+\theta_3}{\phi_0}{\phi_2}+\olag{\theta_3}{\phi_0}{\phi_3}+C(\phi_0)\,, \label{astardef} \\[0.3cm]
A_{\triangle}(\phi_1,\phi_2,\phi_3\,;\,\theta_1,\theta_3)&=\lag{\theta_1}{\phi_2}{\phi_3}+\olag{\theta_1+\theta_3}{\phi_1}{\phi_3}+\lag{\theta_3}{\phi_1}{\phi_2}\,, \label{atridef}
\end{align}
where $C(\phi_i)$ is the leading $O(\hbar^{-1})$ asymptotic of the Boltzmann weight $S(\spin_i)$, given by
\begin{align}
C(\phi_i)=-\left(\frac{1}{rN}\frac{2\phi_i-\frac{r\pi}{2}}{1-\nomec'}\right)^2\,,\qquad 0\leq \re(4rN(1-\nomec')\phi_i+\pi r\nomec')\leq \frac{2\pi}{N}\,,
\end{align}
with period $\phi_i\rightarrow\phi_i+\frac{\pi }{2rN^2(1-\nomec')}$.

The functions $\lag{\theta}{\phi_i}{\phi_j}$ and $\olag{\theta}{\phi_i}{\phi_j}$ are the leading asymptotics of the edge Boltzmann weights \eqref{bwdef_W_lens}, given by
\begin{align}
\begin{split}
&\ds\lag{\theta}{\phi_i}{\phi_j}=\frac{\theta}{\pi}\left(C(\phi_i)+C(\phi_j)\right) \\[0.3cm]
&\ds\hspace{1cm}+\frac{\ii}{rN^2(1-\nomec')}\left\{\int^{\phi_i+\phi_j}_{\frac{r\pi}{2}}dz\log\frac{\vth_3(\theta+z\,|\,r\nomec')}{\vth_3(\theta-z\,|\,r\nomec')}+\int^{\phi_i-\phi_j}_{0}dz\log\frac{\vth_2(\theta+z\,|\,r\nomec')}{\vth_2(\theta-z\,|\,r\nomec')}\right\},
\end{split}
\label{lagdef}
\end{align}
and
\begin{align}
\olag{\theta}{\phi_i}{\phi_j}=\lag{\eta'-\theta}{\phi_i}{\phi_j}\,.
\label{olagdef}
\end{align}
Note that the factorisation \eqref{rfactorisation} has been used to conveniently combine the leading asymptotics of the factor $\rfac{\alpha_1}{\alpha_3}$ in \eqref{bwdef_R}, with the leading asymptotics of the edge Boltzmann weights \eqref{bwdef_W_lens}.

The function \eqref{lagdef} satisfies
\begin{align}
\lag{\theta}{\phi_i}{\phi_j}=\lag{\theta}{\phi_j}{\phi_i}\,,\qquad \lag{-\theta}{\phi_i}{\phi_j}=-\lag{\theta}{\phi_i}{\phi_j}\,,
\end{align}
which are classical manifestations of the symmetries of the Boltzmann weights \eqref{lensBSsym}.

From \eqref{lagdef}, and \eqref{olagdef}, the saddle point equation \eqref{saddle_eq} is explicitly written as
\begin{align}
\left(\leg{\eta'-\theta_1}{\phi_0}{\phi_1}+\leg{\theta_1+\theta_3}{\phi_0}{\phi_2}+\leg{\eta'-\theta_3}{\phi_0}{\phi_3}\right)_{\phi_0=\phi^*}=0\,,
\label{Q43leg}
\end{align}
where
\begin{align}
\leg{\theta}{\phi_i}{\phi_j}=\log\frac{\vth_3(\theta+(\phi_i+\phi_j)\,|\,r\nomec')}{\vth_3(\theta-(\phi_i+\phi_j)\,|\,r\nomec')}+\log\frac{\vth_2(\theta+(\phi_i-\phi_j)\,|\,r\nomec')}{\vth_2(\theta-(\phi_i-\phi_j)\,|\,r\nomec')}\,.
\end{align}

The asymptotics of the star-triangle relation \eqref{STRlim} are governed by solutions of \eqref{Q43leg}, and particularly the classical star-triangle relation \eqref{cSTR}, defined by \eqref{astardef}, and \eqref{atridef}, must hold on solutions $\phi^*$ of \eqref{Q43leg}.  Interestingly, the equation \eqref{Q43leg} appears in the literature as the ``three-leg form'' of $Q4$, the latter being the top level classical discrete integrable equation in a particular classification by Adler, Bobenko, Suris (ABS) \cite{ABS}, based on the concept of ``3D-consistency''. The appearance of $Q4$ in \eqref{Q43leg} is not unexpected, since the equations of the ABS classification have been seen to arise also in the quasi-classical limit of various other solutions of the star-triangle relation \cite{Bazhanov:2007mh,Bazhanov:2010kz,Bazhanov:2016ajm}.  Particularly, the results of the classification of \cite{ABS} suggest that $Q4$ is the only equation that can be expected to appear as the saddle point equation \eqref{Q43leg} at the elliptic level.

\subsection{\texorpdfstring{Subleading order ($O(\hbar^{0})$) expansion:}{Subleading order O(hbar**0) expansion}}

Next consider the subleading order $O(\hbar^0)$ expansion of the star-triangle relation \eqref{STRlim}, where there is now additional dependence on the discrete spin variables $m_i$, and also the discrete variables $n_i$, which enter through the redefinition \eqref{ndef}.

In the following let $\sspin{x}_i$ denote a triplet of variables
\begin{align}
\label{discspindef}
\sspin{x}_i=(\phi_i,m_i,n_i)\,,
\end{align}
where the components are $\phi_i\in\mathbb{R}$, $m_i\in\{0,1,\ldots,r-1\}$, $n_i\in\{0,1,\ldots,N-1\}$.

Define also functions $P_j(\theta\,|\,\sspin{x})$, and $F_j(\theta\,|\,\sspin{x})$ as the following products of Jacobi theta functions
\begin{align}
P_j(\theta\,|\,\sspin{x})&=\!\prod_{k=1}^{rN} \left(\frac{\vartheta_j\left(\frac{\pi n}{N}+\frac{\phi+\theta}{rN}+\frac{\pi(2k-1)}{2rN}\,\Big|\,\frac{\nomec'}{N}\right)}{\vartheta_j\left(\frac{\pi n}{N}+\frac{\phi-\theta}{rN}+\frac{\pi(2k-1)}{2rN}\,\Big|\,\frac{\nomec'}{N}\right)}\right)^{\frac{rN-2(k+m)+1}{2rN}}\prod_{k=1}^{|m|}\left(\frac{\vartheta_j\left(\frac{\pi n}{N}+\frac{\phi+\theta}{rN}-\frac{\sgn(m)\pi(2k-1)}{2rN}\,\Big|\,\frac{\nomec'}{N}\right)}{\vartheta_j\left(\frac{\pi n}{N}+\frac{\phi-\theta}{rN}-\frac{\sgn(m)\pi(2k-1)}{2rN}\,\Big|\,\frac{\nomec'}{N}\right)}\right)^{\sgn(m)}\hspace{-0.4cm}, \\
F_j(\theta\,|\,\sspin{x})&=\prod_{k=1}^{rN}\left(\ovth_j\left(\frac{\pi n}{N}+\frac{\phi+\theta}{rN}+\frac{\pi(2k-1)}{2rN}\,\Big|\,\frac{\nomec'}{N}\right)\,\ovth_j\left(\frac{\pi n}{N}+\frac{\phi-\theta}{rN}+\frac{\pi(2k-1)}{2rN}\,\Big|\,\frac{\nomec'}{N}\right)\right)^{\frac{-1}{2rN}}\hspace{-0.2cm}.
\end{align}
Both of these functions are even functions of the spin $\sspin{x}$ in \eqref{discspindef}, satisfying
\begin{align}
\label{PFsym}
P_j(\theta\,|\,\sspin{x})=P_j(\theta\,|\,-\sspin{x})\,,\qquad F_j(\theta\,|\,\sspin{x})=F_j(\theta\,|\,-\sspin{x})
\end{align}
where $-\sspin{x}=(-\phi,-m,-n)$.

Then the subleading asymptotics of the star-triangle relation \eqref{STRlim} may be written as
\begin{align}
\exp\left(B_{\medstar}(\sspin{x}_0,\sspin{x}_1,\sspin{x}_2,\sspin{x}_3,\theta_1,\theta_3)\right)&=Q(\theta_1\,|\,\sspin{x}_0,\sspin{x}_1)\,P(\theta_1+\theta_3\,|\,\sspin{x}_0,\sspin{x}_2)\,Q(\theta_3\,|\,\sspin{x}_0,\sspin{x}_3)\,\Sd(\sspin{x}_0) \label{bstardef} \,, \\[0.1cm]
\exp\left(B_{\triangle}(\sspin{x}_1,\sspin{x}_2,\sspin{x}_3,\theta_1,\theta_3)\right)&=P(\theta_1\,|\,\sspin{x}_2,\sspin{x}_3)\,Q(\theta_1+\theta_3\,|\,\sspin{x}_1,\sspin{x}_3)\,P(\theta_3\,|\,\sspin{x}_1,\sspin{x}_2)\,R \label{btridef}\,,
\end{align}
where 
\begin{gather}
\label{Pdef}
P(\theta\,|\,\sspin{x}_i,\sspin{x}_j)=P_4(\theta\,|\,\sspin{x}_i+\sspin{x}_j)\,P_1(\theta\,|\,\sspin{x}_i-\sspin{x}_j)\,, \\[0.1cm]
\label{Qdef}
Q(\theta\,|\,\sspin{x}_i,\sspin{x}_j)=P_4(\eta'-\theta\,|\,\sspin{x}_i+\sspin{x}_j)\,P_1(\eta'-\theta\,|\,\sspin{x}_i-\sspin{x}_j)\,F_4(\eta'-\theta\,|\,\sspin{x}_i+\sspin{x}_j)\,F_1(\eta'-\theta\,|\,\sspin{x}_i-\sspin{x}_j) \,,
\end{gather}
are the subleading asymptotics of the edge Boltzmann weights \eqref{bwdef_W_lens} (up to exponential factors that have no overall contribution to the star-triangle relation \eqref{STRlim}).

Due to \eqref{PFsym}, these functions are symmetric in the exchange of $\sspin{x}_i$ and $\sspin{x}_j$, satisfying
\begin{align}
\label{PQsym}
P(\theta\,|\,\sspin{x}_i,\sspin{x}_j)=P(\theta\,|\,\sspin{x}_j,\sspin{x}_i)\,,\qquad Q(\theta\,|\,\sspin{x}_i,\sspin{x}_j)=Q(\theta\,|\,\sspin{x}_j,\sspin{x}_i)\,.
\end{align}

The factor $S(\sspin{x})$ in \eqref{bstardef} is given by
\begin{align}
\label{Sdef}
S(\sspin{x})=\frac{\EXP^{\frac{\pi\ii\nomec'}{8N}}}{\sqrt{2\pi rN}}\;\vartheta_4\left(\frac{2\phi}{rN}+\frac{2\pi n}{N}-\frac{2\pi m}{rN}\,\Big|\,\frac{\nomec'}{N}\right)\,,
\end{align}
and the factor $R$ in \eqref{btridef} is given by
\begin{align}
\label{Rdef}
R=\EXP^{\frac{\pi\ii \nomec'r}{8}}\;\frac{K(\theta_1)K(\theta_3)}{K(\theta_1+\theta_3)}\,,
\end{align}
where
\begin{align}
K(\theta)=\prod_{k=1}^{rN}\, \ovth_1\left(\frac{2\theta}{rN}-\frac{\pi k}{rN}\,\Big|\,\frac{\nomec'}{N}\right)^{\frac{1}{2}-\frac{k}{rN}}\,.
\end{align}
Up to exponential factors (which have no overall contribution to the star-triangle relation \eqref{STRlim}), \eqref{Sdef}, and \eqref{Rdef}, correspond to the subleading asymptotics of the vertex Boltzmann weight \eqref{bwdef_S}, and the normalisation factor \eqref{bwdef_R}, respectively.  Both \eqref{Sdef}, and \eqref{Rdef}, can be derived from the asymptotics of the elliptic gamma function \eqref{key_equation}, for example.

At this order of the expansion, the normalisation factor in \eqref{dSTR} that involves a second derivative of the action also needs to be calculated.  This essentially involves determining a useful expression for the factor $\partial^2 A_{\medstar}(\phi_0\,;\,\phi_1,\phi_2,\phi_3\,;\,\theta_1,\theta_3)/\partial \phi_0^2|_{\phi_0=\phi_*}$, in terms of Jacobi theta functions (or other known functions).  This is non-trivial to compute directly.

One possible method that may be used to find a useful expression for this factor, is to utilise Equation \eqref{dSTR} for $r=N=1$.  Defining 
\begin{align}
\begin{split}
\left. A_{\medstar}(\phi_0\,;\,\phi_1,\phi_2,\phi_3\,;\,\theta_1,\theta_3)\right|_{r=N=1}&:=A^{(1)}_{\medstar}(\phi_0\,;\,\phi_1,\phi_2,\phi_3\,;\,\theta_1,\theta_3)\,, \\[0.1cm]
\left. B_{\medstar}(\phi_0\,;\,\phi_1,\phi_2,\phi_3\,;\,\theta_1,\theta_3)\right|_{r=N=1}&:=B^{(1)}_{\medstar}(\phi_0\,;\,\phi_1,\phi_2,\phi_3\,;\,\theta_1,\theta_3)\,, \\[0.1cm]
\left. B_{\triangle}(\phi_1,\phi_2,\phi_3\,;\,\theta_1,\theta_3)\right|_{r=N=1}&:=B^{(1)}_{\triangle}(\phi_1,\phi_2,\phi_3\,;\,\theta_1,\theta_3)\,,
\end{split}
\end{align}
for $r=N=1$, \eqref{dSTR} may be written as
\begin{align}
\begin{split}
\log\left( \frac{1}{2\pi}\frac{\partial^2 A^{(1)}_{\medstar}(\phi_0\,;\,\phi_1,\phi_2,\phi_3\,;\,\theta_1,\theta_3)}{\partial \phi_0^2} \right)^{-\frac{1}{2}}_{\phi_0=\phi_*} \hspace{-0.45cm}
+B^{(1)}_{\medstar}(\phi_0\,;\,\phi_1,\phi_2,\phi_3\,;\,\theta_1,\theta_3) =B^{(1)}_{\triangle}(\phi_1,\phi_2,\phi_3\,;\,\theta_1,\theta_3)  \,.
\label{sum_saddle_1}
\end{split}
\end{align}
Now note that the difference in the leading asymptotic term $A_\medstar$ for general $r,N\geq 1$, and the leading term $A^1_\medstar$ for $r=N=1$, is a rescaling of $\nomec'$:
\begin{align}
\label{astarn1}
A_{\medstar}(\phi_0\,;\,\phi_1,\phi_2,\phi_3\,;\,\theta_1,\theta_3\,;\,\nomec')=A^{(1)}_{\medstar}\left(\phi_0\,;\,\phi_1,\phi_2,\phi_3\,;\,\theta_1,\theta_3\,;\,r\nomec'\right)\,,
\end{align}
where the dependence on the parameter $\nomec'$, has now been included as the last argument of $A_\medstar$, and $A^{(1)}_\medstar$.  This then implies that a saddle point $\phi_*$ for general $r,N\geq 1$, and a saddle point $\phi^{(1)}_*$ for $r=N=1$, are related by 
\begin{align}
\left. \phi_*\right|_{\nomec=\nomec'}=\left.\phi^{(1)}_*\right|_{\nomec=r\nomec'}\,.
\end{align}
For all cases considered here, the saddle point is in fact independent of $\tau$, and consequently the above saddle points are equivalent.

The general expression for the second derivative evaluated at the saddle point may then be written as
\begin{align}
&\left. \frac{1}{2\pi}\frac{\partial^2 A_{\medstar}(\phi_0\,;\,\phi_1,\phi_2,\phi_3\,;\,\theta_1,\theta_3)}{\partial \phi_0^2}\right|_{\phi_0=\phi_*} =\, \frac{\exp\left(2B^{(1)}_{\medstar}(\phi_0\,;\,\phi_1,\phi_2,\phi_3\,;\,\theta_1,\theta_3\,;\,r\nomec')\right) }
{\exp\left(2B^{(1)}_{\triangle}(\phi_1,\phi_2,\phi_3\,;\,\theta_1,\theta_3\,;\,r\nomec'\right) }\,.
\label{Ad}
\end{align}
Since $\Phi^{(0)}(z,m)=1$ for $r=N=1$, there will be no contribution from any $P_j(\theta\,|\,\sspin{x})$ factors.  Then from the above expressions \eqref{bstardef}, \eqref{btridef}, evaluated at $r=N=1$, the second derivative may be written explicitly as
\begin{align}
\label{seconddfinal}
\left.\frac{\partial^2 A_{\medstar}(\phi_0\,;\,\phi_1,\phi_2,\phi_3\,;\,\theta_1,\theta_3)}{\partial \phi_0^2}\right|_{\phi_0=\phi_*}  = \frac{D(\theta_1\,|\,\phi_*,\phi_1)\,D(\theta_3\,|\,\phi_*,\phi_3)}{D(\theta_1+\theta_3\,|\,\phi_1,\phi_3)}\,,
\end{align}
where
\begin{align}
D(\theta\,|\,\phi_i,\phi_j)=\frac{\ovth_1(2\theta\,|\,r\tau')\,\ovth_4(2\phi_i\,|\,r\tau')\,\ovth_4(2\phi_j\,|\,r\tau')}{\ovth_4(\phi_i+\phi_j\pm\theta\,|\,r\tau')\,\ovth_1(\phi_i-\phi_j\pm\theta\,|\,r\tau')}\,,
\end{align}
and $\pm$ represents a product of two factors with $+$ and $-$ respectively.

Equation \eqref{seconddfinal} is the final expression for the second derivative factor that will be used in the following section, where it will contribute to the $R$ factor of the star-triangle relation.  Note also that similar expressions involving the second derivative of a three-leg form, were previously derived in the consideration of the quasi-classical limit of various different solutions of the star-triangle relation for the case $r=1$ \cite{Bazhanov:2016ajm}.  In fact the expression \eqref{seconddfinal} could be derived from the formulas in Appendix B of \cite{Bazhanov:2016ajm} after using a simple change of variables, due to the property \eqref{astarn1}.

\section{Discrete spin solutions of the star-triangle relation}\label{sec:discrete}

The expressions of the subleading $O(\hbar^{0})$ asymptotics derived in the previous section will be used to write equation \eqref{dSTR}, in the desired form of the star-triangle relation \eqref{STReq}.   Such a star-triangle relation is for an integrable spin model of statistical mechanics, with discrete spins taking values in $\mathbb{Z}_{rN}$.  It will be seen that different solutions of \eqref{Q43leg} correspond to some important integrable models.  However there are additional variables $\phi_0,\phi_1,\phi_2,\phi_3$, which also are parameters of the model that need to be considered, and furthermore $\phi_0$ is required to satisfy the saddle point equation \eqref{Q43leg}.  This makes the general form of the Boltzmann weights quite complicated, except for some special cases.

In this section to simplify notations the primed variables will be written as non-primed variables, so that 
\begin{align}
\label{finalcrossing}
\eta=\frac{\pi}{2}\,,
\end{align}
and $\tau$ is an independent parameter.

Define also the discrete spin variables $\dspin_i$ taking values in $\mathbb{Z}_{rN}$, in terms of the discrete spin components $n_i$, and $m_i$, as
\begin{align}
\label{finalspin}
\dspin_i=rn_i-m_i\,.
\end{align}

Now let $P(\theta\,|\,\phi_i,\phi_j)$ and $Q(\theta\,|\,\phi_i,\phi_j)$ denote the functions \eqref{Pdef}, \eqref{Qdef}, for the case of $m_i=m_j=n_i=n_j=0$, {\it i.e.},
\begin{align}
\label{P0}
P(\theta\,|\,\phi_i,\phi_j):=P(\theta\,|\,(\phi_i,0,0),(\phi_j,0,0))\,, \\
Q(\theta\,|\,\phi_i,\phi_j):=Q(\theta\,|\,(\phi_i,0,0),(\phi_j,0,0))\,.
\end{align}
Then the functions \eqref{Pdef}, \eqref{Qdef} of the previous section can be written in the form
\begin{align}
\label{Pdef2}
P(\theta\,|\,\sspin{x}_i,\sspin{x}_j)&=P(\theta\,|\,\phi_i,\phi_j)\,\bw{\theta}{(\phi_i,a_i)}{(\phi_j,a_j)}\,, \\[0.1cm]
\label{Qdef2}
Q(\theta\,|\,\sspin{x}_i,\sspin{x}_j)&=Q(\theta\,|\,\phi_i,\phi_j)\,\olbw{\theta}{(\phi_i,a_i)}{(\phi_j,a_j)}\,,
\end{align}
where the edge Boltzmann weights in \eqref{Pdef2}, \eqref{Qdef2} are given by
\begin{align}
\label{dw1}
\bw{\theta}{(\phi_i,a_i)}{(\phi_j,a_j)}&=w_{4,3}(\theta\,|\,\phi_i+\phi_j,a_i+a_j)\, w_{1,2}(\theta\,|\,\phi_i-\phi_j,a_i-a_j)\,,
\end{align}
and
\begin{align}
\label{dw2}
\olbw{\theta}{(\phi_i,a_i)}{(\phi_j,a_j)}&=\bw{\eta-\theta}{(\phi_i,a_i)}{(\phi_j,a_j)}\,,
\end{align}
where
\begin{align}
\begin{split}
w_{j_1,j_2}(\theta\,|\,\phi,a)&=\left(\frac{\ovth_{j_2}(\phi+\theta\,|\,r\nomec)}{\ovth_{j_2}(\phi-\theta\,|\,r\nomec)}\right)^{\frac{a}{rN}}
\;\times\left\{\begin{array}{ll}
\ds\prod_{k=1}^{a}\,\frac{\ovth_{j_1}\left(\frac{\phi-\theta}{rN}+\frac{\pi(2k-1)}{2rN}\,\Big|\,\frac{\nomec}{N}\right)}{\ovth_{j_1}\left(\frac{\phi+\theta}{rN}+\frac{\pi(2k-1)}{2rN}\,\Big|\,\frac{\nomec}{N}\right)}& \quad a\geq1\,,\\[0.6cm]
\ds\prod_{k=1}^{-a}\,\frac{\ovth_{j_1}\left(\frac{\phi+\theta}{rN}-\frac{\pi(2k-1)}{2rN}\,\Big|\,\frac{\nomec}{N}\right)}{\ovth_{j_1}\left(\frac{\phi-\theta}{rN}-\frac{\pi(2k-1)}{2rN}\,\Big|\,\frac{\nomec}{N}\right)}& \quad a<1\,.
\end{array}\right.
\end{split}
\end{align}

The function $w_{j_1,j_2}(\theta\,|\,\phi,a)$ satisfies the periodicity with respect to $rN$
\begin{align}
\label{BWsym}
w_{j_1,j_2}(\theta\,|\,\phi,a)=w_{j_1,j_2}(\theta\,|\,\phi,a+rN)\,.
\end{align}
Consequently the edge Boltzmann weights \eqref{dw1}, \eqref{dw2}, satisfy a similar periodicity.  The edge Boltzmann weights also satisfy the symmetry
\begin{align}
\label{BWchiral}
\bw{\theta}{(\phi_i,a_i)}{(\phi_j,a_j)}=\bw{\theta}{(\phi_j,a_j)}{(\phi_i,a_i)}\,,
\end{align}
and
\begin{align}
\label{BWnorm}
\bw{\theta}{(\phi_i,0)}{(\phi_j,0)}=1\,.
\end{align}
Because of the symmetry \eqref{BWchiral}, except for some special choices of the $\phi_i$, $\phi_j$, satisfying \eqref{Q43leg}, the Boltzmann weights will generally be chiral with respect to the discrete spins $a_i$, $a_j$.

The corresponding vertex Boltzmann weight is given by
\begin{align}
\label{Sdef2}
S(\phi,a)=\frac{\EXP^{\frac{\pi\ii\nomec}{8N}}}{\sqrt{rN}}\,\vth_4\left(\frac{2\pi a}{rN}+\frac{2\phi}{rN}\,\Big|\,\frac{\nomec}{N}\right)\,,
\end{align}
and satisfies a similar a periodicity to the edge Boltzmann weights
\begin{align}
S\left(\phi,a+\frac{rN}{2}\right)=S(\phi,a)\,.
\end{align}

The normalisation factor $R$ consists of a combination of the remaining factors coming from \eqref{Rdef}, \eqref{seconddfinal}, \eqref{Pdef2}, \eqref{Qdef2}, and depends on the variables $\theta_1,\theta_3$, and $\phi_0,\phi_1,\phi_2,\phi_3$.  The $R$ factor is defined in terms of \eqref{P0}, and the following factor
\begin{align}
\label{smallrdef}
\begin{split}
r(\theta\,|\,\phi_i,\phi_j)&=\frac{\EXP^{\frac{\pi\ii r\nomec}{8}}\,\vth_1(2\theta\,|\,r\nomec)}{Q(\theta\,|\,\phi_i,\phi_j)}\left(\frac{\ovth_4(2\phi_x\,|\,r\nomec)\,\ovth_4(2\phi_y\,|\,r\nomec)}{\ovth_4(\phi_x+\phi_y\pm\theta\,|\,r\nomec)\,\ovth_1(\phi_x-\phi_y\pm\theta\,|\,r\nomec)\,}\right)^{\frac{1}{2}}\prod_{k=1}^{rN}\ovth_1\left(\frac{2\theta-\pi k}{rN}\,\Big|\,\frac{\nomec}{N}\right)^{\frac{-k}{rN}}\hspace{-0.2cm},
\end{split}
\end{align}
as
\begin{align}
\label{Rdef2}
R=\frac{r(\theta_1\,|\,\phi_0,\phi_1)\,r(\theta_3\,|\,\phi_0,\phi_3)}{r(\theta_1+\theta_3\,|\,\phi_1,\phi_3)}\frac{P(\theta_1\,|\,\phi_2,\phi_3)\,P(\theta_3\,|\,\phi_1,\phi_2)}{P(\theta_1+\theta_3\,|\,\phi_0,\phi_2)}\,.
\end{align}
Note that the ordering of two spins $\phi_i$, $\phi_j$, doesn't matter in arguments of $r(\theta\,|\,\phi_i,\phi_j)$, and $P(\theta\,|\,\phi_i,\phi_j)$, which follows from the definition \eqref{smallrdef}, and the symmetries \eqref{PQsym}.

Then the Boltzmann weights \eqref{dw1}, \eqref{dw2}, \eqref{Sdef2}, \eqref{Rdef2}, satisfy the star-triangle relation:
\begin{align}
\begin{split}
\sum_{a_0=0}^{rN-1}\!S(\phi_*,a_0)&\olbw{\theta_1}{(\phi_*,a_0)}{(\phi_1,a_1)}\bw{\theta_1+\theta_3}{(\phi_*,a_0)}{(\phi_2,a_2)}\olbw{\theta_3}{(\phi_0,a_0)}{(\phi_3,a_3)}  \\[-0.1cm]
=R\,&\bw{\theta_1}{(\phi_2,a_2)}{(\phi_3,a_3)}\olbw{\theta_1+\theta_3}{(\phi_1,a_1)}{(\phi_3,a_3)}\bw{\theta_3}{(\phi_1,a_1)}{(\phi_2,a_2)} ,
\end{split}
\label{STRdef2}
\end{align}
where $\phi_*$ is a solution to the saddle point equation \eqref{Q43leg}.  This is the final expression for the star-triangle relation corresponding to an integrable $\mathbb{Z}_{rN}$-state lattice model.  For the case $r=1$, and $\theta_i\to\theta_i/2$, $\phi_i\to\phi_i/2$, this is equivalent to the star-triangle relation for $\mathbb{Z}_N$ models obtained by Bazhanov and Sergeev, in Section 4.2 of \cite{Bazhanov:2010kz}.  Note also that the latter star-triangle relation (for $r=1$) \cite{Bazhanov:2010kz}, may be obtained with the change of variables $(rN,\tau)\to(N,\tau N)$, in \eqref{STRdef2}.  The latter connection between the two star-triangle relations is rather surprising, considering the difference in the subleading order asymptotics of the lens elliptic gamma function \eqref{slead}, for the respective cases of $r=1$, and $r>1$.  Consequently the root of unity limit (up to $O(1)$) of the star-triangle relation \eqref{STReq}, is effectively independent of $r$, and thus the cases of the star-triangle relation \eqref{STRdef2} studied here, are equivalent to the well known cases (Kashiwara-Miwa and chiral Potts models) that were also previously studied in \cite{Bazhanov:2010kz}.

In the star-triangle relation \eqref{STRdef2}, $\phi_*$ is a solution of the saddle point equation \eqref{Q43leg}, and hence is taken to be a fixed value and is not a free parameter.  Due to the symmetry of the problem (to avoid inhomogeneities in the spin system) the same comment applies to $\phi_1, \phi_2, \phi_3$.  The role of the $\phi_i$ also changes now, where the $\phi_i$ originally comes from the redefinition of spins \eqref{COV} in the model \eqref{STReq}, the choice of the $\phi_*,\phi_1,\phi_2,\phi_3$, as well as $\theta_1,\theta_3$, now will coincide with the rapidity parameterisation in \eqref{STRdef2}.

The saddle point equation \eqref{Q43leg} in principle can be solved exactly for $\phi_*$, in terms of $\phi_1,\phi_2,\phi_3$, and this solution will be given in terms of Jacobi (or Weierstrass) elliptic functions and their inverses.  This is a direct consequence of the ``affine-linear'' property of the $Q4$ equation \cite{ABS}.  Substituting this value of $\phi_*$ into the discrete spin star-triangle relation \eqref{STRdef2} results in a general elliptic solution of the star-triangle relation, with $a_0, a_1, a_2, a_3\in \mathbb{Z}_{rN}$ defined in \eqref{finalspin}, now playing the role of the spins of the model.  However for a general choice of $\phi_*,\phi_1,\phi_2,\phi_3$ satisfying the saddle point equation \eqref{Q43leg}, the Equations \eqref{dw1}, and \eqref{dw2}, will give a complicated expression for up to six independent Boltzmann weights satisfying \eqref{STRdef2}, and an interpretation as an integrable model of statistical mechanics is not immediately obvious.  Nonetheless, as seen for the $r=1$ case \cite{Bazhanov:2010kz}, there are some special symmetric solutions of \eqref{Q43leg}, where the resulting Boltzmann weights \eqref{dw1}, and \eqref{dw2}, satisfying \eqref{STRdef2}, correspond to well-known integrable lattice models of statistical mechanics.

In the following subsections, the known solutions \cite{Bazhanov:2010kz} of \eqref{Q43leg} will be considered, along with some details not included in \cite{Bazhanov:2010kz}, and also some new aspects.  First, in Section \ref{sec:kmmodel} the solution of \eqref{Q43leg} corresponding to the Kashiwara-Miwa model is considered.  The Boltzmann weights obtained directly from the normalisation used in \eqref{BWnorm} are given, along with the explicit connection to the usual convention used for the Kashiwara-Miwa model \cite{Bax02rip,Bazhanov:2010kz}.  In Section \ref{sec:triglim}, the trigonometric limit of \eqref{STRdef2} will be considered, which results in the general expression for Boltzmann weights satisfying \eqref{STRdef2}, that are given in terms of the $\cos$ and $\sin$ functions.  For this case, the saddle-point equation for \eqref{STRdef2} reduces from $Q4$ in \eqref{Q43leg}, to $Q3_{(\delta=0)}$ in \eqref{Q33leg}.  In Section \ref{sec:fzmodel}, it is shown that the Kashiwara-Miwa solution of the saddle-point equation \eqref{Q43leg} used in Section \ref{sec:kmmodel}, also solves $Q3_{(\delta=0)}$ in \eqref{Q33leg}, which results in the Boltzmann weights of the Fateev-Zamolodchikov model.  In Section \ref{sec:cpmodel}, the chiral Potts solution \cite{Bazhanov:2010kz} of $Q3_{(\delta=0)}$ is considered.  Particularly, it is shown here that with the usual form of Boltzmann weights, the chiral Potts spectral curve is the unique solution to the trigonometric saddle point equation \eqref{Q33leg}, at least under a certain parametrization of the spectral parameter.  This provides new insight into the appearance of the chiral Potts curve in the rapidity parameterisation of the chiral Potts model, as the unique solution of the saddle point equation corresponding to the classical integrable lattice equation $Q3_{(\delta=0)}$.

\subsection{Kashiwara-Miwa model}\label{sec:kmmodel}

A special case of the solution of \eqref{Q43leg} is found for the values $\phi_0,\phi_1,\phi_2,\phi_3=\frac{\pi}{2}(\zeta+\nu)$, where $\zeta\in\mathbb{Z}$, and $\nu\in\{0,\tfrac{1}{2}\}$.  This corresponds to the case of the Kashiwara-Miwa model \cite{Kashiwara:1986tu}, with the edge Boltzmann weights given by
\begin{align}
\label{KMbw}
\begin{split}
\bw{\theta}{a_i}{a_j}&=w_4(\theta\,|\,\pi(\zeta+\nu),a_i+a_j)\,w_1(\theta\,|\,0,a_i-a_j)\,, \\[0.1cm]
\olbw{\theta}{a_i}{a_j}&=\bw{\eta-\theta}{a_i}{a_j}\,,
\end{split}
\end{align}
where
\begin{align}
w_j(\theta\,|\,\phi,a)=\left\{\begin{array}{ll}
\ds\prod_{k=1}^{a}\,\frac{\ovth_j\left(\frac{\phi-\theta}{rN}-\frac{\pi(2k-1)}{2rN}\,\Big|\,\frac{\nomec}{N}\right)}{\ovth_j\left(\frac{\phi+\theta}{rN}-\frac{\pi(2k-1)}{2rN}\,\Big|\,\frac{\nomec}{N}\right)}& \quad a\geq1\,, \\[0.6cm]
\ds\prod_{k=1}^{-a}\,\frac{\ovth_j\left(\frac{\phi+\theta}{rN}+\frac{\pi(2k-1)}{2rN}\,\Big|\,\frac{\nomec}{N}\right)}{\ovth_j\left(\frac{\phi-\theta}{rN}+\frac{\pi(2k-1)}{2rN}\,\Big|\,\frac{\nomec}{N}\right)}& \quad a<1\,,
\end{array}\right.
\end{align}
and the vertex Boltzmann weight given by
\begin{align}
\label{KMs}
S(a)=\frac{1}{\sqrt{rN}}\,\vth_4\left(\frac{2\pi a}{rN}+\frac{\pi(\zeta+\nu)}{rN}\,\Big|\,\frac{\nomec}{N}\right)\,.
\end{align}
The crossing parameter is defined in \eqref{finalcrossing}.

The Boltzmann weights above satisfy the periodicity with respect to $rN$
\begin{align}
\label{KMper}
\bw{\theta}{a_i}{a_j}=\bw{\theta}{a_i+rN}{a_j}=\bw{\theta}{a_i}{a_j+rN}\,,\qquad S(a+rN)=S(a)\,,
\end{align}
while the edge Boltzmann weights also satisfy the symmetry
\begin{align}
\label{KMsym}
\bw{\theta}{a_i}{a_j}=\bw{\theta}{a_j}{a_i}\,,
\end{align}
and similarly for $\olbw{\theta}{a_i}{a_j}$.

The edge Boltzmann weights also satisfy
\begin{align}
\label{KMnorm}
\bw{\theta}{0}{0}=\olbw{\theta}{0}{0}=1\,.
\end{align}

For this choice of the $\phi_0,\phi_1,\phi_2,\phi_3$, many theta functions cancel in the normalisation factor \eqref{Rdef2}, which may be written in the form
\begin{align}
\label{KMr}
R(\theta_1,\theta_3)=\frac{r(\theta_1)\,r(\theta_3)}{r(\theta_1+\theta_3)}\,.
\end{align}
where
\begin{align}
\begin{split}
r(\theta)&=\vth_4(\pi\nu\,|\,r\nomec)\,\ovth_2(\theta\,|\,r\nomec)\,\ovth_3(\theta+\pi\nu\,|\,r\nomec)\prod_{k=1}^{2\,\sgn(\zeta)(\zeta+\nu)}\frac{\ovth_4\left(\frac{\theta}{rN}-\frac{\pi k}{2rN}\,\Big|\,\frac{\nomec}{N}\right)}{\ovth_4\left(\frac{\theta}{rN}+\frac{\pi (k-1)}{2rN}\,\Big|\,\frac{\nomec}{N}\right)} \\[0.2cm]
&\times\prod_{k=0}^{rN-1}\left(\ovth_2\left(\frac{\theta}{rN}+\frac{\pi k}{2rN}\,\Big|\,\frac{\nomec}{N}\right)\,\ovth_3\left(\frac{\theta}{rN}+\frac{\pi k}{2rN}\,\Big|\,\frac{\nomec}{N}\right)\right)^{-1}\,.
\end{split}
\end{align}
Note that this $r(\theta)$ also depends on the two parameters $\zeta$, and $\nu$.

The above Equations \eqref{KMbw}, \eqref{KMs}, \eqref{KMr}, define a solution of the star-triangle relation
\begin{align}
\begin{split}
&\sum_{a_0=0}^{rN-1}S(a_0)\,\olbw{\theta_1}{a_0}{a_1}\,\bw{\theta_1+\theta_3}{a_0}{a_2}\,\olbw{\theta_3}{a_0}{a_3}  \\[0.1cm]
&\qquad=R(\theta_1,\theta_3)\,\bw{\theta_1}{a_2}{a_3}\,\olbw{\theta_1+\theta_3}{a_1}{a_3}\,\bw{\theta_3}{a_1}{a_2} \,.
\end{split}
\end{align}

This star-triangle relation depends on a number of different variables, these are the three independent spins $a_1,a_2,a_3$, as well as two independent spectral variables $\theta_1,\theta_3$, and it also depends on the values of the additional parameters $\tau$, $\zeta$, $\nu$, and the product $rN$.

The edge and vertex Boltzmann weights \eqref{KMbw}, \eqref{KMs}, have a similar form to the usual Boltzmann weights for the Kashiwara-Miwa model, while the normalisation factor $R$ in \eqref{KMr} has a rather different form from the usual normalisation factor $R$, for example given in \cite{Bax02rip,Bazhanov:2010kz}.  The reason for this is some differences in the definitions of the Boltzmann weights, including the different condition on Boltzmann weights \eqref{KMnorm}, which is inherited from the original model \eqref{STReq}.  By manipulating the products in \eqref{KMr}, it is straightforward to relate the Boltzmann weights to the usual form.

First the $r(\theta)$ above can be written into the equivalent form
\begin{align}
\begin{split}
r(\theta)&=\prod_{k=1}^{\lfloor \frac{rN}{2}\rfloor}\frac{\ovth_1\left(\frac{\pi(2k-1)}{2rN}+\frac{\theta}{rN}\,\Big|\,\frac{\nomec}{N}\right)}{\ovth_1\left(\frac{\pi k}{rN}-\frac{\theta}{rN}\,\Big|\,\frac{\nomec}{N}\right)}\,\prod_{k=1}^{\lfloor \frac{rN}{2}-\nu\rfloor}\frac{\ovth_4\left(\frac{\pi(2k-1+2\nu)}{2rN}+\frac{\theta}{rN}\,\Big|\,\frac{\nomec}{N}\right)}{\ovth_4\left(\frac{\pi(k+\nu)}{rN}-\frac{\theta}{rN}\,\Big|\,\frac{\nomec}{N}\right)} \\[0.2cm]
&\times\vth_4(\pi\nu\,|\,r\nomec)\,\left(\frac{\ovth_4\left(\frac{\theta}{rN}-\frac{\pi}{2rN}\,\Big|\,\frac{\nomec}{N}\right)}{\ovth_4\left(\frac{\theta}{rN}\,\Big|\,\frac{\nomec}{N}\right)}\right)^{2\nu}\,\prod_{k=1}^{2\,\sgn(\zeta)(\zeta+\nu)}\frac{\ovth_4\left(\frac{\theta}{rN}-\frac{\pi k}{2rN}\,\Big|\,\frac{\nomec}{N}\right)}{\ovth_4\left(\frac{\theta}{rN}+\frac{\pi(k-1)}{2rN}\,\Big|\,\frac{\nomec}{N}\right)}\,.
\end{split}
\label{Rkm2}
\end{align}
The products in the first line are already equivalent to the usual $r(\theta)$ factor of the model.  The $\vth_4(\pi\nu\,|\,r\nomec)$ factor should be moved to the definition of $S(a)$.  For the remaining terms, first note that
\begin{align}
\begin{split}
w_4(\theta\,|\,\pi(\zeta+\nu),a_i+a_j)=t_4(\theta\,|\,a_i+a_j+\zeta)\prod_{k=1}^{|\zeta|}\,\frac{\ovth_4\left(\frac{\sgn(zt)\pi\nu+\theta}{rN}+\frac{\pi(2k-1)}{2rN}\,\Big|\,\frac{\nomec}{N}\right)}{\ovth_4\left(\frac{\sgn(zt)\pi\nu-\theta}{rN}+\frac{\pi(2k-1)}{2rN}\,\Big|\,\frac{\nomec}{N}\right)}\,,
\end{split}
\end{align}
where
\begin{align}
t_j(\theta\,|\,a)=\left\{\begin{array}{ll}
\ds\prod_{k=1}^{a}\,\frac{\ovth_j\left(\frac{\pi\nu-\theta}{rN}+\frac{\pi(2k-1)}{2rN}\,\Big|\,\frac{\nomec}{N}\right)}{\ovth_j\left(\frac{\pi\nu+\theta}{rN}+\frac{\pi(2k-1)}{2rN}\,\Big|\,\frac{\nomec}{N}\right)}& \quad a\geq1\,, \\[0.6cm]
\ds\prod_{k=1}^{-a}\,\frac{\ovth_j\left(\frac{\pi\nu+\theta}{rN}-\frac{\pi(2k-1)}{2rN}\,\Big|\,\frac{\nomec}{N}\right)}{\ovth_j\left(\frac{\pi\nu-\theta}{rN}-\frac{\pi(2k-1)}{2rN}\,\Big|\,\frac{\nomec}{N}\right)}& \quad a<1\,.
\end{array}\right.
\end{align}
Then the remaining products of theta functions in \eqref{Rkm2}, exactly cancel with the extra terms that come from rewriting the edge Boltzmann weights of \eqref{KMbw}, in the form
\begin{align}
\bw{\theta}{a_i}{a_j}=t_4(\theta\,|\,a_i+a_j+\zeta)\,w_1(\theta\,|\,0,a_i-a_j)\,.
\end{align}
This edge Boltzmann weight satisfies the same properties \eqref{KMper}, and \eqref{KMsym}, while instead of \eqref{KMnorm}, this Boltzmann weight now satisfies
\begin{align}
\bw{\theta}{0}{0}=t_4(\theta\,|\,\zeta)\,.
\end{align}
This is the form of the edge Boltzmann weight for the Kashiwara-Miwa model given in \cite{Bax02rip,Bazhanov:2010kz}.


\subsection{Trigonometric limit}\label{sec:triglim}

Consider next the trigonometric limit\footnote{At the level of the lens elliptic gamma function solution of the Yang-Baxter equation \eqref{bwdef_W_lens} (in the original variables), this more closely resembles a hyperbolic limit \cite{GahramanovKels}.}, when $\nomec\rightarrow\ii\infty$, of \eqref{STRdef2}.  First, in this limit the saddle point equation of motion \eqref{Q43leg} becomes
\begin{align}
\left(\leg{\eta-\theta_1}{\phi_0}{\phi_1}+\leg{\theta_1+\theta_3}{\phi_0}{\phi_2}+\leg{\eta-\theta_3}{\phi_0}{\phi_3}\right)_{\phi_0=\phi^*}=0\,,
\label{Q33leg}
\end{align}
where
\begin{align}
\leg{\theta}{\phi_i}{\phi_j}=\log\frac{\cos(\theta+(\phi_i-\phi_j))}{\cos(\theta-(\phi_i-\phi_j))}\,.
\end{align}
The saddle point equation \eqref{Q33leg} may be identified with the ``three-leg form'' of classical discrete integrable equation known as $Q3_{(\delta=0)}$ \cite{ABS}.

The edge Boltzmann weights \eqref{dw1}, \eqref{dw2} in the trigonometric limit are
\begin{align}
\label{trigBW1}
\bw{\theta}{(\phi_i,a_i)}{(\phi_j,a_j)}=\left(\frac{\cos(\phi_i-\phi_j+\theta)}{\cos(\phi_i-\phi_j-\theta)}\right)^{\frac{a_i-a_j}{rN}}\!\!\times\left\{\begin{array}{ll}
\ds\prod_{k=1}^{a_i-a_j}\,\frac{\sin\left(\frac{\phi_i-\phi_j-\theta}{rN}+\frac{\pi(2k-1)}{2rN}\right)}{\sin\left(\frac{\phi_i-\phi_j+\theta}{rN}+\frac{\pi(2k-1)}{2rN}\right)}&\;\; a_i-a_j\geq1\,,\\[0.6cm]
\ds\prod_{k=1}^{a_j-a_i}\,\frac{\sin\left(\frac{\phi_i-\phi_j+\theta}{rN}-\frac{\pi(2k-1)}{2rN}\right)}{\sin\left(\frac{\phi_i-\phi_j-\theta}{rN}-\frac{\pi(2k-1)}{2rN}\right)}&\;\; a_i-a_j<1\,,
\end{array}\right.
\end{align}
and
\begin{align} 
\label{trigBW2}
\olbw{\theta}{(\phi_i,a_i)}{(\phi_j,a_j)}=\bw{\eta-\theta}{(\phi_i,a_i)}{(\phi_j,a_j)}\,,
\end{align}
where the crossing parameter is defined in \eqref{finalcrossing}.
As in the elliptic case \eqref{BWsym}, the Boltzmann weights satisfy the symmetries
\begin{align}
&\bw{\theta}{(\phi_i,a_i+rN)}{(\phi_j,a_j)}=\bw{\theta}{(\phi_i,a_i)}{(\phi_j,a_j+rN)}=\bw{\theta}{(\phi_i,a_i)}{(\phi_j,a_j)}\,,\\
&\bw{\theta}{(\phi_i,a_i)}{(\phi_j,a_j)}=\bw{\theta}{(\phi_j,a_j)}{(\phi_i,a_i)} \label{trigsym}\,,
\end{align}
and similarly for $\olbw{\theta}{(\phi_i,a_i)}{(\phi_j,a_j)}$.

The $S$ factor \eqref{Sdef2} becomes
\begin{align}
S(a_i)=1\,,
\end{align}
and the normalisation factor \eqref{Rdef2} becomes
\begin{align}
\label{trigR}
R=\frac{r(\theta_1\,|\,\phi_0-\phi_1)\,r(\theta_3\,|\,\phi_3-\phi_0)}{r(\theta_1+\theta_3\,|\,\phi_3-\phi_1)}\frac{P(\theta_1\,|\,\phi_3-\phi_2)\,P(\theta_3\,|\,\phi_1-\phi_2)}{P(\theta_1+\theta_3\,|\,\phi_0,\phi_2)}\,,
\end{align}
where
\begin{align}
r(\theta\,|\,\phi)=\frac{(rN)^{\frac{1}{2}}\,\sin(2\theta)^{\frac{1}{2}}}{\left(2\,\sin(\phi\pm\theta)\right)^{\frac{rN-1}{2rN}}\,P(\frac{\pi}{2}-\theta\,|\,\phi)} \,\prod_{k=1}^{rN}\sin\left(\frac{2\theta-\pi k}{rN}\right)^{\frac{rN-2k}{2rN}}\,,
\end{align}
and
\begin{align}
P(\theta\,|\,\phi)=\left(\frac{\cos(\phi+\theta)}{\cos(\phi-\theta)}\right)^{\frac{rN+1}{2rN}}\,\prod_{k=1}^{rN}\,\left(\frac{\sin\left(\frac{\phi+\theta}{rN}+\frac{\pi(2k-1)}{2rN}\right)}{\sin\left(\frac{\phi-\theta}{rN}+\frac{\pi(2k-1)}{2rN}\right)}\right)^{-\frac{k}{rN}}\,.
\end{align}
Some cancellation has been used between identical exponential factors that appear in both \eqref{Sdef2}, and \eqref{Rdef2}.

The functions \eqref{trigBW1}, \eqref{trigBW2}, \eqref{trigR}, define a solution of the star-triangle relation \eqref{STRdef2}.  Note that in the above trigonometric limit, the $S$ factor is $S(a_i)=1$, and the star-triangle relation will only depend on differences of the spin variable components, in the form $\phi_i-\phi_j$, and $a_i-a_j$.  Thus the particular choices of $\phi_0,\phi_1,\phi_2,\phi_3$ here satisfying the saddle point equation \eqref{Q33leg}, result in solutions of the star-triangle relation with $\mathbb{Z}_{rN}$ symmetry.  For the elliptic case of the previous section, this symmetry is broken.\footnote{Note that also some continuous spin solutions of STR are known \cite{Spiridonov:2010em,GahramanovKels} with broken $\mathbb{Z}_{rN}$ symmetry, and which are not elliptic.}

\subsection{Fateev-Zamolodchikov model} \label{sec:fzmodel}

The same solution $\phi_0=\phi_1=\phi_2=\phi_3=\frac{\pi}{2}(\zeta+\nu)$, $\zeta\in\mathbb{Z}$, $\nu\in\{0,\frac{1}{2}\}$, used to obtain the Kashiwara-Miwa model in Section \ref{sec:kmmodel}, can be used as a solution to \eqref{Q33leg}.  However in this case there is no dependence on the additional parameters $\zeta$, and $\nu$, as they can be simply absorbed into a redefinition of the discrete spins $a_i$.  As expected, this solution of the saddle point equation corresponds to the Fateev-Zamolodchikov model \cite{Fateev:1982wi}, with Boltzmann weights given by
\begin{align}
\label{bwfz}
\bw{\theta}{a_i}{a_j}=\prod_{k=1}^{|a_i-a_j|}\frac{\sin\left(\frac{\theta}{rN}-\frac{\pi(2k-1)}{2rN}\right)}{\sin\left(\frac{-\theta}{rN}-\frac{\pi(2k-1)}{2rN}\right)},\qquad \olbw{\theta}{a_i}{a_j}=\bw{\frac{\pi}{2}-\theta}{a_i}{a_j}\,,
\end{align}
and
\begin{align}
\label{srfz}
S(a_i)=1\,,\qquad R=\frac{r(\theta_1)\,r(\theta_3)}{r(\theta_1+\theta_3)}\,,
\end{align}
where
\begin{align}
\label{rfacfz}
r(\theta)=\sqrt{rN}\,\prod_{k=1}^{\lfloor \frac{rN}{2}\rfloor}\frac{\sin\left(\frac{\theta}{rN}+\frac{\pi(2k-1)}{2rN}\right)}{\sin\left(\frac{\theta}{rN}+\frac{\pi k}{rN}\right)}\,.
\end{align}
The factorisation of $R$ in \eqref{rfacfz} agrees with the expression obtained in \cite{Bax02rip}.

The Boltzmann weights are obviously symmetric, satisfying
\begin{align}
\bw{\theta}{a_i}{a_j}=\bw{\theta}{a_j}{a_i}\,,\qquad\olbw{\theta}{a_i}{a_j}=\olbw{\theta}{a_j}{a_i}\,,
\end{align}
and are periodic by a shift of the spins by $rN$
\begin{align}
\bw{\theta}{a_i}{a_j}=\bw{\theta}{a_i+rN}{a_j}=\bw{\theta}{a_i}{a_j+rN}\,,\\[0.1cm] 
\olbw{\theta}{a_i}{a_j}=\olbw{\theta}{a_i+rN}{a_j}=\olbw{\theta}{a_i}{a_j+rN}\,.
\end{align}

The above Boltzmann weights \eqref{bwfz}, and \eqref{srfz}, define a solution of the star-triangle relation
\begin{align}
\bs
\sum_{a_0=1}^{rN-1}&\olbw{\theta_1}{a_1}{a_0}\,\bw{\theta_1+\theta_3}{a_2}{a_0}\,\olbw{\theta_3}{a_3}{a_0} \\[0.1cm]
&=R\,\bw{\theta_1}{a_2}{a_3}\,\olbw{\theta_1+\theta_3}{a_1}{a_3}\,\bw{\theta_3}{a_1}{a_2}\,.
\es
\end{align}

\subsection{Chiral Potts model} \label{sec:cpmodel}

\subsubsection{Reparametrization}

For the cases of the Kashiwara-Miwa and Fateev-Zamolodchikov models, the parameters $\theta_1$ and $\theta_3$ 
are identified directly with the spectral parameters of the model.
We can instead try to identify spectral parameters from 
some combination of $\theta_1$ and $\theta_3$, as well as 
$\phi_0, \dots, \phi_3$.

An example of this is the following parametrization introduced by Bazhanov and Sergeev
\cite{Bazhanov:2010kz}, 
where we have six 
parameters\footnote{The rapidity variables labelled $x_r$, and $y_r$, are not related to the integer parameter $r$, defined in \eqref{rparam}.} $x_p,y_p,x_q,y_q,x_r,y_r$, which are
defined up to overall multiplication:
 \begin{align}
\bs
&\EXP^{\frac{2\ii\theta_1}{rN}}=\sqrt{\frac{x_{r}y_{r}}{x_{q}y_{q}}}\;, \quad \EXP^{\frac{2\ii\theta_3}{rN}}=\sqrt{\frac{x_{q}y_{q}}{x_{p}y_{p}}} \;,\\
& \EXP^{\frac{2\ii(\phi_1-\phi_0)}{rN}}=\sqrt{\frac{y_{q}x_{r}}{x_{q}y_{r}}} \;,\ \quad  \EXP^{\frac{2\ii(\phi_2-\phi_0)}{rN}}=\omega^{\frac{1}{2}}\sqrt{\frac{x_{p}x_{r}}{y_{p}y_{r}}}\;, \quad  \EXP^{\frac{2\ii(\phi_3-\phi_0)}{rN}}=\sqrt{\frac{x_{p}y_{q}}{y_{p}x_{q}}} \;,
\label{CP_parametrization}
\es
\end{align}
where we used $\omega=\EXP^{2\pi\ii/rN}$.

In this parametrization, we find that the saddle point equation \eqref{Q33leg} may be written as
\be
X_1 Y_1  (X_2-X_3+Y_2-Y_3)
+X_2 Y_2 (X_3-X_1+Y_3-Y_1)
+X_3 Y_3  (X_1-X_2+Y_1-Y_2)
=0 \;,
\label{simple}
\ee
where we defined 
\be
X_i:=x_i^{rN}\;, \quad Y_i:=y_i^{rN} \;.
\label{XY}
\ee
Note that the final equation is manifestly symmetric in the
cyclic exchange of indices $1,2,3$.

In terms of the parametrization introduced in \eqref{CP_parametrization},
the Boltzmann weights (\eqref{trigBW1} and \eqref{trigBW2}) to be evaluated at the saddle points are given by 
\begin{align}
\label{cpBW1}
W_{pq}(a_i,a_j)=\left(\frac{y_{p}^{rN}-x_{q}^{rN}}{y_{q}^{rN}-x_{p}^{rN}}\right)^{\frac{a_i-a_j}{rN}}\!\!\times\left\{
\begin{array}{cl}
\ds\prod^{a_i-a_j}_{k=1}\frac{y_{q}-\omega^k x_{p}}{y_{p}-\omega^kx_{q}}\,, & \quad a_i-a_j\geq1\,,\\[0.6cm]
\ds\prod^{a_j-a_i-1}_{k=0}\frac{x_{q}-\omega^k y_{p}}{x_{p}-\omega^ky_{q}}\,, &\quad a_i-a_j<1\,,
\end{array}\right.
\end{align}
and
\begin{align}
\label{cpBW2}
\olW_{pq}(a_i,a_j)=\left(\frac{y_{p}^{rN}-y_{q}^{rN}}{x_{q}^{rN}-x_{p}^{rN}}\right)^{\frac{a_i-a_j}{rN}}
\!\!\times\left\{
\begin{array}{cl}
\ds\prod^{a_i-a_j}_{k=1}\frac{\omega x_{p}-\omega^kx_{q}}{y_{q}-\omega^ky_{p}}\,, & \quad a_i-a_j\geq1\,,\\[0.6cm]
\ds\prod^{a_j-a_i-1}_{k=0}\frac{ y_{p}-\omega^ky_{q}}{x_{q}-\omega^{k+1}x_{p}}\,, &\quad a_i-a_j<1\,.
\end{array}\right.
\end{align}
These Boltzmann weights are chiral upon the exchange of spins $a_i$, $a_j$, {\it i.e.}
\begin{align}
W_{pq}(a_i,a_j)\neq W_{pq}(a_j,a_i)\,,\qquad \olW_{pq}(a_i,a_j)\neq \olW_{pq}(a_j,a_i)\,.
\end{align}
This is expected since the Boltzmann weights \eqref{trigBW1}, \eqref{trigBW2} are only symmetric upon exchange of the pairs $(\phi_i,a_i)$ and $(\phi_j,a_j)$, as in \eqref{trigsym}.

The Boltzmann weight \eqref{cpBW1} also satisfies the usual periodicity with respect to $rN$
\begin{align}
W_{pq}(a_i+rN,a_j)=W_{pq}(a_i,a_j+rN)=W_{pq}(a_i,a_j)\,,
\end{align}
and similar for \eqref{cpBW2}.

Note that these Boltzmann weights define a solution of the star-triangle relation, with the only restriction on the variables $x_p,y_p,x_q,y_q,x_r,y_r$, being that they satisfy the saddle point equation \eqref{simple}.  To have the usual interpretation as integrable model of statistical mechanics, it remains to determine a solution of \eqref{simple}, such that pairs $(x_p,y_p)$, $(x_q,y_q)$, $(x_r,y_r)$, may be interpreted as rapidity variables of the corresponding model.

\subsubsection{Chiral Potts curve}

A special solution of the saddle point equation \eqref{simple},
is given when the pairs $(x_p,y_p)$ satisfy the usual spectral curve of the chiral Potts model:
\begin{align}
x^{rN}_{p}+y_{p}^{rN}=k(1+x^{rN}_{p}y^{rN}_{p}) \;.
\label{CP_curve_xy}
\end{align}
We can then regard the points $(x_{p},y_{p})$ on this curve
as rapidity variables. 

The Boltzmann weights \eqref{cpBW1} and \eqref{cpBW2} may be written respectively as
\begin{align}
\label{cpBW12}
W_{pq}(a_i,a_j)=\left(\frac{\mu_{p}}{\mu_{q}}\right)^{a_i-a_j}\!\!\times\left\{
\begin{array}{cl}
\ds\prod^{a_i-a_j}_{k=1}\frac{y_{q}-\omega^kx_{p}}{y_{p}-\omega^kx_{q}}\,, & \quad a_i-a_j\geq1\,,\\[0.6cm]
\ds\prod^{a_j-a_i-1}_{k=0}\frac{x_{q}-\omega^ky_{p}}{x_{p}-\omega^ky_{q}}\,, &\quad a_i-a_j<1\,,
\end{array}\right.
\end{align}
and
\begin{align}
\label{cpBW22}
\olW_{pq}(a_i,a_j)=\left(\mu_{p}\mu_{q}\right)^{a_i-a_j}\!\!\times\left\{
\begin{array}{cl}
\ds\prod^{a_i-a_j}_{k=1}\frac{\omega x_{p}-\omega^kx_{q}}{y_{q}-\omega^ky_{p}}\,, & \quad a_i-a_j\geq1\,,\\[0.6cm]
\ds\prod^{a_j-a_i-1}_{k=0}\frac{ y_{p}-\omega^ky_{q}}{x_{q}-\omega^{k+1}x_{p}}\,, &\quad a_i-a_j<1\,.
\end{array}\right.
\end{align}
where we have introduced $\mu_{p}$ defined by 
\begin{align}
kx^{rN}_{p}=1-k'\mu_{p}^{-rN}\,,\quad ky^{rN}_{p}=1-k'\mu_{p}^{rN}\,,
\end{align}
with $k^2+k'^2=1$.  The point $(x_p, y_p, \mu_p)$ defines a point on the spectral curve.

 These  above Boltzmann weights then exactly coincide with the Boltzmann weights of the chiral Potts model \cite{AuYang:1987zc,Baxter:1987eq}. The Boltzmann weights \eqref{cpBW12} and \eqref{cpBW22} do not satisfy a rapidity difference property, and hence depend on the values of the two rapidity variables, labelled by $p$, and $q$, independently.
 
The $R$ factor \eqref{trigR} may be written as
\begin{align}
\label{cpR}
R_{pqr}=\frac{f_{qr}f_{rp}}{f_{pq}}\,,\qquad \left(f_{pq}\right)^{rN}=\prod_{k=1}^{rN-1}\left(\frac{\mu_{q}(x_{q}-\omega^ky_{p})}{\mu_{p}(x_{p}-\omega^ky_{q})}\frac{(1-\omega^k)(x_{p}y_{p}-\omega^kx_{q}y_{q})}{(x_{p}-\omega^kx_{q})(y_{p}-\omega^ky_{q})}\right)^{k}\,.
\end{align}
The $f_{pq}$ is written here as a product of two fractions, the fraction on the left is the contribution coming from factors of $P(\theta\,|\,\phi_i-\phi_j)$ in \eqref{trigR}, and the fraction on the right is the contribution coming from factors of $r(\theta\,|\,\phi_i-\phi_j)$ in \eqref{trigR}.

The Boltzmann weights \eqref{cpBW12}, \eqref{cpBW22}, \eqref{cpR}, then satisfy the star-triangle relations
\begin{align}
\begin{split}
&\sum_{a_0=0}^{rN-1}\olW_{qr}(a_1,a_0)\,W_{pr}(a_2,a_0)\,\olW_{pq}(a_0,a_3) =R_{pqr}\,W_{qr}(a_2,a_3)\,\olW_{pr}(a_1,a_3)\,W_{pq}(a_2,a_1) \,, \\
&\sum_{a_0=0}^{rN-1}\olW_{qr}(a_0,a_1)\,W_{pr}(a_0,a_2)\,\olW_{pq}(a_3,a_0) =R_{pqr}\,W_{qr}(a_3,a_2)\,\olW_{pr}(a_3,a_1)\,W_{pq}(a_1,a_2) \,.
\end{split}
\end{align}

\subsubsection{Uniqueness: homogeneous case}

In the analysis of the previous subsection, the Boltzmann weights
\eqref{cpBW1} and \eqref{cpBW2} already take the same form as those for the chiral Potts model,
even before choosing a specific saddle point as given by the chiral Potts curve \eqref{CP_curve_xy}.
It is then a natural question to ask if 
we can find more general solutions to the saddle point equation,
whose associated spectral curve is  different from \eqref{CP_curve_xy}.

Let us consider the possibility that the 
all the points $(x_{p}, y_{p})$ are located on the 
same spectral curve $\mathcal{C}$, of the form 
\be
\mathcal{C}: Y=F(X) \;,
\ee
where the capitalized variables are introduced in \eqref{XY}.\footnote{Written in this form, $F(X)$ in general is a multi-valued function of $X$,
however this fact will not affect the following analysis.}

This assumption has an immediate consequence.
Since all the points $(x_{p}, y_{p})$ are located on the 
same spectral curve $\mathcal{C}$, we have
$Y_i=F(X_i)$ with a single function $F(X)$ independent of the index $i$.

An immediate consequence for this is that the saddle point equation \eqref{simple} now 
gives a functional equation
\be
\bs
&X_1 F(X_1)  (X_2-X_3+F(X_2)-F(X_3))
+X_2 F(X_2) (X_3-X_1+F(X_3)-F(X_1))\\
&\qquad+X_3 F(X_3)  (X_1-X_2+F(X_1)-F(X_2))=0 \,.
\label{F_XYZ}
\es
\ee

Since the equation is preserved the the shift of $F(X)$ by a constant,
we can assume $F(X=0)=0$.\footnote{We here implicitly assumed that the origin $X=0$ is contained in the spectral curve.
The argument here, however, can be applied to other points, say $X=1$, with only minor modifications.}
Then we have, by choosing $X_3=0$,
\be
\bs
\frac{F(X_1)-F(X_2)}{X_1-X_2}=-\frac{F(X_1) F(X_2)}{X_1 X_2}  \,.
\label{F_constraint}
\es
\ee
By taking the limit $X_2\to X_1=X$ we obtain
\be
F'(X)=- \frac{F(X)^2}{X^2} \,.
\ee
to give
\be
Y=F(X)=\frac{X}{CX-1}-D \,,
\ee
for some constants $C, D$ (here we recovered the constant part $F(0)=-D$). 
This gives the spectral curve 
\be
CXY+D=X+Y \;.
\ee

If $C, D\ne 0$, 
by simultaneous rescaling of $X$ and $Y$ we can set $C=D=k\ne 0$,
to obtain the curve for the chiral Potts model \eqref{CP_curve_xy}.
If $C=0$ or $D=0$ we obtain simpler curves
\be
x^{rN}+y^{rN}=D \;, \quad \textrm{or} \quad Cx^{rN} y^{rN}=x^{rN}+y^{rN} \,,
\label{CD0}
\ee
which can be thought of as degenerations of the chiral Potts curves \eqref{CP_curve_xy}. 

This completes the proof that under the assumptions above,
the chiral Potts model, as described by the curve \eqref{CP_curve_xy},
is the unique possibility.

\subsubsection{Uniqueness: inhomogeneous case}

In light of the uniqueness argument of the previous section,
the natural question is to explore more possibilities by relaxing some conditions.

One such possibility is to consider a different spectral curve for each rapidity line,
so that we have
\be
Y_1=F(X_1) \;, \quad Y_2=G(X_2) \;, \quad Y_3=H(X_3) \,.
\ee
for three independent functions $F(X), G(X), H(X)$.
We then obtain a functional equation
\be
\bs
&X_1 F(X_1)  (X_2-X_3+G(X_2)-H(X_3))
+X_2 G(X_2) (X_3-X_1+H(X_3)-F(X_1))\\
&\qquad+X_3 H(X_3)  (X_1-X_2+F(X_1)-G(X_2))=0 \,.
\label{FGH}
\es
\ee

As before, we can choose $F(X=0)=0$,
by shifting both $F(X)$ and $G(X)$ by the same constant.
Then by choosing $X_1=0$ we have
\be
\bs
&
 \frac{G(X_2) H(X_3)}{X_2 X_3}=-\frac{G(X_2)  - H(X_3)}{X_2 -X_3}  \,.
\label{G_constraint}
\es
\ee
Consistency of this equation as $X_2\to X_3$ requires\footnote{This argument does not apply 
when the spectral curve for the rapidity parameters
$(x_2, y_2)$ and $(x_3, y_3)$ 
collapses to a single point, so that we have 
\be
Y_1=F(X_1) \;,\quad Y_2=Y_3=\beta\;, \quad X_2=X_3=\alpha  \,.
\ee
Indeed, the saddle point equation is automatically satisfied for any choice of the function $F(X)$ in this case.
Unfortunately, the corresponding model is problematic since the Boltzmann weight associated with the crossing of rapidities lines for $(x_2, y_2)$ and $(x_3, y_3)$ then diverges.
}
$G(X)=H(X)$. Then \eqref{G_constraint}
is the same constraint as before (see \eqref{F_constraint}),
and hence determines $G(X)$ to be of the chiral Potts form:
$G(X)=X/(C X-1)-D$. By plugging this into \eqref{FGH},
and using \eqref{G_constraint}, and $H(X)=G(X)$, 
we have
\be
F(X_1)= \frac{X_1 (X_2 G(X_2)-X_3 G(X_3) ) }
{X_1(X_2+G(X_2)-X_3-G(X_3))-(X_2 G(X_2)-X_3 G(X_3) ) }\,.
\ee
The right hand side is independent of $X_2, X_3$ only 
if we have $D=0$, in which case we obtain
\be
F(X)=\frac{X}{CX-1}=G(X)  \,.
\ee
We therefore find that the chiral Potts model is again the 
unique possibility, even under this relaxed condition.

\section{Comments on Gauge Theory Interpretation}\label{sec:gauge}

As commented in introduction, the lens elliptic gamma function solution of the STR we started with, 
naturally arises from the lens index \cite{Benini:2011nc}
of four-dimensional $\mathcal{N}=1$ quiver gauge theories \cite{Yamazaki:2013nra}.
It is then natural to ask for the interpretation of the results above in the language of supersymmetric gauge theories.
While we leave the detailed analysis for future work, let us below make some preliminary comments.

\subsection{\texorpdfstring{Geometry of $S^1 \times S^3/\mathbb{Z}_r$}{Geometry of S1*S3/Zr}}

The lens index is the supersymmetric partition function on the geometry
$S^1\times S^3/\mathbb{Z}_r$, with the complex structure parametrised by $\p, \q$ as \cite{Closset:2013vra}
\begin{align}
(z_1, z_2 ) &\sim (\p^2 z_1, \q^2 z_2) \;,\label{eq.quotient_pq}\\
(z_1, z_2) &\sim (e^{\frac{2\pi \ii}{r}} z_1, e^{-\frac{2\pi \ii}{r}} z_2) \;,
\label{eq.quotient_r}
\end{align}
where \eqref{eq.quotient_pq} defines $S^3\times S^1$ and 
\eqref{eq.quotient_r} defines its quotient by $\mathbb{Z}_r$.

Indeed, we can apply a coordinate transformation
\begin{align}
\begin{split}
&z_1=e^{2\pi \ii \tau_1x }  \cos\frac{\theta}{2} e^{\ii \varphi}\;, \quad z_2=e^{2\pi \ii \tau_2 x}\sin\frac{\theta}{2} e^{\ii \chi} \;,\\
&x\sim x+1 \;, \quad 0\le \theta \le \pi \;, \quad \varphi\sim \varphi+2 \pi \;, \quad \chi \sim \chi+2\pi \;.
\label{eq.quotient}
\end{split}
\end{align}
Then $x$ is the coordinate for $S^1$, while $(\theta, \varphi, \chi)$ the coordinates for $S^3$,
expressed as a $T^2$-fibration (parametrized by $\varphi, \chi$) over an interval $[0, \pi]$ (parametrized by $\theta$) 
with one of the one-cycles of the fiber degenerating at the two endpoints.
This $S^3$ is an ellipsoid, which at position $x$ of $S^1$ is given by 
\begin{align}
(e^{2\pi \textrm{Im} \tau_1 x} )^2 |z_1|^2 +(e^{2\pi \textrm{Im}  \tau_2 x})^2 |z_2|^2=1 \;.
\end{align}
The $\mathbb{Z}_r$-quotient
\eqref{eq.quotient_r} acts as $2\pi/r$-rotation along the $S^1$-fiber for the Hopf fiberation of $S^3$.

We need  $|\p|, |\q|<1$ in order for the geometry defined by equations \eqref{eq.quotient} to be compact.\footnote{Note that the first equation can equivalently be written as 
$(z_1, z_2 ) \sim (\p^{-1} z_1, \q^{-1} z_2)$, implying the geometrical symmetry
as $(\p,\q) \leftrightarrow (\p^{-1}, \q^{-1})$. This means we can equivalently take
$|\p|, |\q|>1$. The superconformal index
indeed has such a symmetry, if the flavor fugacities are simultaneously 
inverted, see \cite{Basar:2014mha}.}
This is also needed for the convergence of the lens elliptic gamma function, 
and hence of the associated lens superconformal index.

\subsection{Saddle point and Gauge/Bethe correspondence}

Let us first consider the un-orbifolded case $r=1$,
and take the limit $\q\to 1$. In our previous notation,
this is to consider the root-of-unity limit \eqref{root_of_unity_limit} with $N=1$.

In this limit, the geometry $S^1\times S^3$ will decompactify
into $\mathbb{T}^2\times \mathbb{C}$;
the identification \eqref{eq.quotient_pq} reduces in the limit to $z_1\to \p z_1$, 
which defines a torus with modulus $2\pi \tau_1$, and $z_2$ is a coordinate of $\mathbb{C}$.
The parameter $\hbar$, when kept finite, has the effect of regularizing the non-compact geometry,
as is similar to the case of the $\Omega$-background \cite{Moore:1997dj,Nekrasov:2002qd}.

One can make this point more precise. 
In the limit $\hbar\to 0^{+}$ the lens index is divergent, as we have seen above:
\begin{align}
I \to \int d\sigma \, \exp{ \left[\frac{1}{\hbar} I^{(-1)}(\sigma) + I^{(0)}(\sigma)
+ \mathcal{O}(\hbar^1)  \right]} \;.
\label{I_expand}
\end{align}
We propose that the leading divergence $I^{(-1)}(\sigma)$ should be
identified with the effective twisted superpotential $\mathcal{W}_{\rm eff}(\sigma)$ of the
four-dimensional $\mathcal{N}=1$ theories on $\mathbb{T}^2$---when compactified, 
we obtain two-dimensional $\mathcal{N}=(2,2)$ theories with infinitely-many Kaluza-Klein (KK) modes, and after integrating out massive matters
we obtain an effective twisted superpotential, a function of the complex scalar in the 
adjoint $\mathcal{N}=(2,2)$ vector multiplet.\footnote{This statement should be a limit of
the proposed factorization of the four-dimensional index \cite{Yoshida:2014qwa,Peelaers:2014ima,Nieri:2015yia,Lodin:2017lrc}.}
A similar analysis was made in \cite{Yamazaki:2012cp} for the three-dimensional $\mathcal{N}=2$ theory on $S^3$ when four-dimensional theory on $S^1\times S^3$ is compactified along $S^1$ (with KK modes along the compactified $S^1$ neglected), and
our discussion here can be thought of an ``elliptic uplift'' of the story there.

We have seen that the leading piece is build out of the function $\Phi^{(-1)}$ defined in \eqref{lead},
which function is written as an integral of the logarithm of the Jacobi theta function:\footnote{Since $r=N=1$
we have $\hat{\nomec}=\nomec+1$ in \eqref{lead}.}
\begin{align}
 \int_0^z du\log\ovth_4\left(u\,|\,\nomec+1\right)=
  \int_0^z du\log\ovth_3\left(u\,|\,\nomec\right)
 \,.
 \label{ell_Li}
\end{align}
This can be rewritten as an infinite sum of the classical dilogarithm function:
\begin{align}
\begin{split}
&
\int_0^z du \left( 
    \ln \prod_{j=1}^{\infty} (1+ \EXP^{2 \ii u} \EXP^{\ii\pi N \tau(2j-1)})(1+\EXP^{-2 \ii u} \EXP^{\ii\pi N\tau(2j-1)})
\right)\,.\\
&\quad 
= \sum_{j=0}^{\infty} 
\left(
\int_0^z \ln(1+\EXP^{\ii u} \EXP^{\ii\pi N\tau(2j+1)}) du
-
\int_0^{-z} \ln(1+\EXP^{\ii u} \EXP^{\ii\pi N\tau(2j+1)}) du
\right)
\\
& \quad =
\ii  \sum_{j=0}^{\infty} 
\left(
 \textrm{Li}_2\left(-\EXP^{2N \ii z} \EXP^{\ii\pi N\tau(2j+1)} \right) 
- \textrm{Li}_2\left(-\EXP^{-2N \ii z}  \EXP^{\ii\pi  N\tau(2j+1)} \right) 
\right) \;,
\end{split}
\end{align}
where we used the integral representation of the classical dilogarithm function
\begin{align}
\ii \, \textrm{Li}_2(-e^{c+ix })
=  \int^x\, du \,  \ln (1+e^{c+iu}) 
\end{align}
for constant $c$. 
Such an infinite sum of the classical dilogarithm function
also appears in the twisted superpotential for a
four-dimensional $\mathcal{N}=1$ chiral multiplet \cite{Nekrasov:2009uh};
the infinite sum in the equation above represents the sum over KK modes when we compactify from 
four dimensions to three dimensions, and the function \eqref{ell_Li} should be 
thought of as elliptic version of the classical dilogarithm function (see also \cite{Closset:2017bse} for recent related discussion).

We can push this correspondence further.
In our analysis of the quasi-classical limit of the STR,
it was crucial to solve the saddle point equation for the leading piece:
\begin{align}
\exp\left(\frac{\partial \mathcal{I}^{(-1)} (\sigma)}{\partial \sigma} \right) 
=\exp\left(\frac{\partial \mathcal{W}_{\rm eff} (\sigma)}{\partial \sigma} \right) =1 \;.
\end{align}

This equation is also the equation determining the vacua of 
the two-dimensional $\mathcal{N}=(2,2)$ theory,
and was studied in the context of 
the so-called Gauge/Bethe correspondence
of Nekrasov and Shatashvili \cite{Nekrasov:2009uh}.
There the saddle point equation of the two-dimensional model is identified with the Bethe 
Ansatz equation of the associated integrable model, and that the two-dimensional theory arises from four-dimensional $\mathcal{N}=1$ theory on $\mathbb{T}^2$ is reflected in the fact that the associated integrable model is governed by 
an elliptic version of the quantum group.

This should be compared with the discussion above (in the context of the Gauge/YBE correspondence \cite{Yamazaki:2012cp,Terashima:2012cx,Yamazaki:2013nra}), where the saddle point equation of the leading part $I^{(-1)}$ is identified with a 
classical discrete integrable equation of \cite{ABS}.
Here we also have a version of the elliptic quantum group---as shown in \cite{Zabrodin:2010qm} (see also \cite{Maruyoshi:2016caf}) the $R$-matrix for our integrable model 
with $r=1$ \cite{Bazhanov:2010kz,Spiridonov:2010em,Yamazaki:2012cp,Terashima:2012cx} arises as the intertwiner for two representations of the Sklyanin algebra $U_{\p,\q}(\mathfrak{sl}_2)$ \cite{Sklyanin:1983ig} (an elliptic algebra associated with the $R$-matrix for the eight-vertex model \cite{Baxter:1971cr}).\footnote{When the gauge groups of four-dimensional quiver gauge theory is a product of $SU(N_c)$ we have 
the algebra $U_{\p,\q}(\mathfrak{sl}_{N_c})$ of \cite{CherednikGeneralized}.}

Indeed, the parallel becomes even more striking once we consider the Gauge/Bethe correspondence for the four-dimensional $\mathcal{N}=2$ theory on the $\Omega$-background \cite{Nekrasov:2009rc}; there we have two equivariant parameters $\epsilon_1, \epsilon_2$ for $U(1)^2$ actions on
$\mathbb{C}^2$, and the Nekrasov-Shatashvili limit is the limit for $\epsilon_2\to 0$ with $\epsilon_1$ kept finite.
This is the same limit studied in this paper if we identify $\tau_1, \tau_2$ with $\epsilon_1, \epsilon_2$.

We can summarize the comparison between Gauge/YBE and Gauge/Bethe correspondence as in Table \ref{table.compare}.

\begin{table}[htbp]
\begin{center}
\caption{Parallel between Gauge/YBE and Gauge/Bethe correspondence.}
\medskip
\label{table.compare}
\begin{tabular}{c|c}
  Gauge/YBE correspondence & Gauge/Bethe correspondence\\
 \hline
 \hline
 4d $\mathcal{N}=1$ theory  & 4d $\mathcal{N}=2$ theory \\ 
   \hline
  superconformal index & Nekrasov partition function \\ 
   \hline
  $(S^3\times S^1)_{\p, \q}$ & $ (\mathbb{C}^2)_{\epsilon_1, \epsilon_2}$\\ 
 \hline
  superconformal index fugacities $(\p, \q)$  & $\Omega$-background parameters $(\epsilon_1, \epsilon_2)$ \\ 
   \hline
  unity limit  $\q\to 1$  & Nekrasov-Shashvili limit $\epsilon_2 \to 0$ \\   
     \hline
 leading piece $I^{(-1)}(\sigma)$ of index &effective twisted superpotential $\mathcal{W}_{\rm eff}(\sigma)$\\ 
      \hline
 saddle point equation & vacuum equation for 2d $\mathcal{N}=(2,2)$ theory \\ 
\hline
  discrete integrable equation ($Q4$) & Bethe Ansatz equation  \\
   \hline
  Sklyanin/Cherednik algebra $U_{\p,\q}(\mathfrak{sl}_{N_c})$  & elliptic quantum group \\
      \hline
root-of-unity limit ($N>1$) & ??? \\ 
    \hline
  $\mathbb{Z}_r$-orbifolding & ??? \\ 
\end{tabular}
\end{center}
\end{table}

There is clearly more need to explore this parallel further, especially for the 
more general cases of $r>1$ and $N>1$ discussed in this paper, 
whose counterpart in the context of the Gauge/Bethe correspondence seems to be unknown.
We hope to return to this exciting topic in the near future.

\section{Conclusion}

\begin{figure}[tbh]
\centering
\begin{tikzpicture}[scale=1]
\draw[white!] (-7,3) circle (0.01pt)
node[above=1pt]{\color{black} Quantum};
\draw[white!] (7,3) circle (0.01pt);
\draw[white!] (-7,0.25) circle (0.01pt)
node[above=1pt]{\color{black} Classical};
\draw[white!] (7,0.25) circle (0.01pt);

\draw[white!] (2.0,1.7) circle (0.01pt)
node[right=1pt]{\color{black}\small \cite{Bazhanov:2016ajm} $\; N\to\infty$};
\draw[white!] (-2.0,1.7) circle (0.01pt)
node[left=1pt]{\color{black}\small  $\q\to\EXP^{\pi\ii/rN}$};
\draw[white!] (0,3) circle (0.01pt)
node[above=1pt]{\color{black} MS}
node[above=30pt]{\color{black}\small  $\;\q\to\EXP^{\pi\ii/rN}$ }
node[below=20pt]{\color{black}\small \cite{Bazhanov:2010kz} $\q\to\EXP^{\pi\ii/N}$ };
\draw[white!] (1.8,3.6) circle (0.01pt)
node[below=5pt]{\color{black}\small \cite{Bazhanov:2010kz} $\;\q\to\EXP^{\pi\ii/N}$};
\draw[white!] (-1.8,3.6) circle (0.01pt)
node[below=5pt]{\color{black}\small $r=1$};
\begin{scope}[>=Latex]
\draw[thick,->>] (0.6,3.4)--(3.4,3.4);
\end{scope}
\draw[thick,<-] (-0.6,3.4)--(-3.3,3.4);
\begin{scope}[>=Latex]
\draw[thick,->>] (-3.6,3.6) .. controls (-1.5,4.2) and (1.5,4.2) .. (3.6,3.6);
\end{scope}
\draw[white!] (-4,3) circle (0.01pt)
node[above=1pt]{\color{black} LEGF};
\draw[white!] (4,3) circle (0.01pt)
node[above=1pt]{\color{black} KM};
\draw[thick,-] (0,3)--(0,2.25);
\begin{scope}[>=Latex]
\draw[thick,->] (0,1.75)--(0,1);
\end{scope}
\draw[white!] (0,0.25) circle (0.01pt)
node[above=1pt]{\color{black} $Q4$};
\begin{scope}[>=Latex]
\draw[thick,->] (-4,3)--(-0.5,1);
\end{scope}
\begin{scope}[>=Latex]
\draw[thick,->] (4,3)--(0.5,1);
\end{scope}

\end{tikzpicture}
\caption{Elliptic quantum/classical integrable models correspondence involving solutions of the star-triangle relation.  Here LEGF, MS, KM, stand for lens elliptic gamma function solution (considered in this paper), master solution \cite{Bazhanov:2010kz}, Kashiwara-Miwa solution \cite{Kashiwara:1986tu}, of star-triangle relations respectively.  Also filled single, and double arrow heads, respectively represent leading ($O(\hbar^{-1})$) and subleading ($O(\hbar^0)$) order quasi-classical limits.}
\label{ellipticfig}
\end{figure}
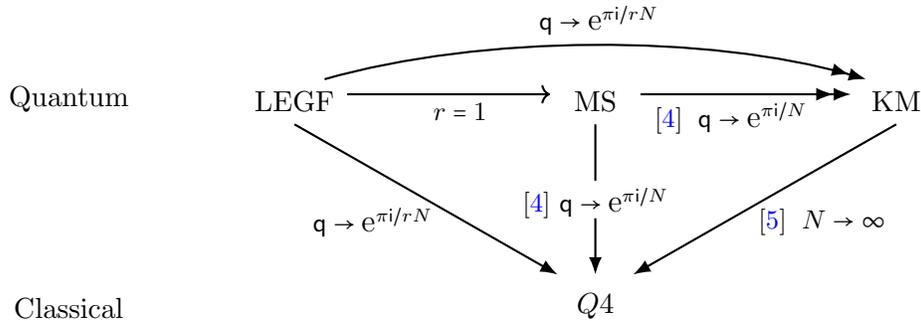

In this paper it is shown how the root of unity limit of the lens elliptic gamma function solution of the star-triangle relation \eqref{STReq}, reduces to the well-known discrete spin solutions of the star-triangle relations, namely the Kashiwara-Miwa model, at the elliptic level, and chiral Potts and Fateev-Zamolodchikov models, at the trigonometric level.  Furthermore, the specific integrable model that is obtained in the root of unity limit, corresponds to a certain solution of the classical integrable lattice equation $Q4$ \cite{ABS}, or in the trigonometric limit, $Q3_{(\delta=0)}$.  This provides an important new example of the recently observed correspondence \cite{Bazhanov:2007mh,Bazhanov:2010kz,Bazhanov:2016ajm,Kels:2017fyt}, between quantum and classical integrable systems that satisfy the Yang-Baxter equation, and the 3D-consistency properties, respectively.  The resulting correspondence between the different quantum and classical integrable models at the elliptic level is summarised in the diagram in Figure \ref{ellipticfig}.

In the quasi-classical limit, the original discrete spins with values in $\mathbb{Z}_r$, and new discrete spins with values in $\mathbb{Z}_N$ (corresponding to a certain periodicity associated with the $2rN$-th root of unity), were shown in Section \ref{sec:discrete} to be effectively described in terms of a single discrete spin, with values in $\mathbb{Z}_{rN}$.  This is consistent with the previous calculations for the $r=1$ case \cite{Bazhanov:2010kz}, which resulted in the same discrete spin integrable models, with $\mathbb{Z}_N$ valued spins.  Consequently the root of unity limits of the $r=1$, and $r>1$ cases respectively, essentially coincide up to a change of $N\to rN$.  This is a rather non-trivial and unexpected connection, particularly considering that the subleading order asymptotics of the lens elliptic gamma function \eqref{slead} for $r=1$ (with $m=0$), and $r>1$, are quite different.    Consequently, the low temperature limit of \eqref{STReq} is essentially independent of $r$, at least up to $O(1)$.  This result, and connection outlined in Figure \ref{ellipticfig}, provides some insight into the properties of the integrable models with continuous and discrete spins that are based on the Boltzmann weights \eqref{bwdef_W_lens}.  For example, it is of interest to determine the quantum group structure that underlies the lens elliptic gamma function solutions of the Yang-Baxter equation \cite{Yamazaki:2013nra,Kels:2015bda}.  The R-matrices for \eqref{bwdef_W_lens}, for the $r=1$, and $r>1$ cases are rather different, and consequently the quantum group structure for general $r>1$, is likely to differ from that of the special case $r=1$ (Sklyanin algebra $U_{\p,\q}(\mathfrak{sl}_2)$ \cite{Sklyanin:1983ig}).  The connection established in Figure \ref{ellipticfig} suggests that whatever algebra is associated to general $r$, say $U_{\p,\q; r}(\mathfrak{sl}_2)$, the two algebras $U_{\p,\q}(\mathfrak{sl}_2)$ and $U_{\p,\q; r}(\mathfrak{sl}_2)$ should coincide in the root of unity limit.  This will be of help in identifying any new algebraic structure, and in the study of its representation theory.  Similarly this connection gives possible insight into the other so far unknown properties of the model, such as diagonalization of transfer matrices, and expressions for observables such as magnetization, where in the root of unity limit they can be expected to reduce to the known quantities for the respective discrete spin models.

In Section \ref{sec:discrete}, it has also been shown that for a particular parametrisation of Boltzmann weights, the chiral Potts curve is the unique restriction on the rapidity variables, required to satisfy the saddle point equation corresponding to $Q3_{(\delta=0)}$.  This gives a new perspective into the appearance of the chiral Potts curve in the rapidity parameterisation of the chiral Potts model, and  
particularly this implies that the chiral Potts model can be expected to be the most general form of the discrete spin star-triangle relation at the hyperbolic level, which possesses $\mathbb{Z}_N$ symmetry.  One thing to note here, is that while the different integrable models of statistical mechanics in Section \ref{sec:discrete}, were shown to correspond to different solutions of classical integrable lattice equations $Q4$, and $Q3_{\delta=0}$,  conversely it is not true that any solution of the latter integrable lattice equations, will correspond to an integrable model of statistical mechanics.  For example, there is a different case of the classical $Q3$ equation, where the parameter is $\delta=1$ (rather than $\delta=0$), however the solutions that were used for the case $\delta=0$ in Sections \ref{sec:fzmodel}, and \ref{sec:cpmodel}, either do not provide a solution for $\delta=1$, or result in singular Boltzmann weights.  It would be interesting to find a solution of $Q3_{(\delta=1)}$ leading to non-singular Boltzmann weights, corresponding to a hyperbolic solution of the star-triangle relation with broken $\mathbb{Z}_N$ symmetry, of which there are no known cases (as far as the authors are aware).  In this direction it could be possible to generalise the uniqueness argument presented in Section \ref{sec:cpmodel}, in order to obtain such a model for the case of $Q3_{(\delta=1)}$.  The same also applies to the elliptic case of $Q4$, if a suitable parameterization of the latter equation can be found, which would indicate whether there are some other elliptic solutions of the star-triangle relation with discrete integer valued spins, apart from the Kashiwara-Miwa model.

Finally in Section \ref{sec:gauge}, details were given on how the results of this paper may be interpreted with respect to the gauge/YBE correspondence \cite{Yamazaki:2013nra}, which provides a relation between the four-dimensional $\mathcal{N}=1$ quiver gauge theory on the lens space, to the two-dimensional integrable model of this paper.  In this context, the root of unity limit of the lens supersymmetric index for the four-dimensional $\mathcal{N}=1$ theory, is seen to correspond to the effective twisted superpotential for a certain two-dimensional $\mathcal{N}=(2,2)$ theory.  However with respect to the results of this paper, there are certain important aspects of the correspondence which still remain to be fully understood, including the gauge theory interpretation for the appearance of the discrete spin star-triangle relations in Section \ref{sec:discrete}, and the relation to the gauge/Bethe correspondence of Nekrasov and Shatashvili, in light of the parallels summarised in Table \ref{table.compare}.

There are several other important directions for future work which are worth mentioning.  For example, a natural next step is to consider the root of unity limit of multi-spin models \cite{Yamazaki:2013nra,Kels:2017toi}, that are a direct generalisation of the star-triangle relation \eqref{STReq}.  For these multi-spin models, even in the simplest $r=1$ case \cite{Bazhanov:2011mz,Bazhanov:2013bh} the corresponding root of unity limit is not yet known.  It is expected that the analysis of these multi-spin cases should follow closely to the analysis of the root of unity limit in Sections \ref{sec:limit}, and \ref{sec:discrete}.  Another possibility is to consider hyperbolic solutions of the Yang-Baxter equation \cite{Bazhanov:2007vg,Spiridonov:2010em,GahramanovKels}, and corresponding multi-spin cases. These cases have an extra complication due to the integration being taken over the entire real line, which would possibly lead to discrete spin models with arbitrary integer valued spins in $\mathbb{Z}$ (instead of $\mathbb{Z}_{rN}$).  Each of the above cases may lead to new integrable models, and importantly provide new examples of a correspondence between the Yang-Baxter equation, and 3D-consistency conditions for multi-component spin variables, where not much is known, particularly at the classical level.   Finally, it is also of interest to determine whether the root of unity limit considered in this paper, has some application in the context of certain integrable spin chains, which were recently shown \cite{Derkachov:2016dhc,Derkachov:2016ucn,Derkachov:2017pvx} to result in sum/integral formulas related to some limits of the star-triangle relation \eqref{STReq}.  We hope to return to each of these interesting topics in our future works.

\section*{Acknowledgements}

The authors initiated this project at the workshop ``Baxter 2015: Exactly Solved Models and Beyond'', in Palm Cove, Australia, and would like to thank the organisers, particularly Vladimir Bazhanov, for their hospitality and for providing financial support to attend the workshop. 
APK is an overseas researcher under Postdoctoral Fellowship of Japan Society for the Promotion of Science (JSPS).
The research of MY in various stages was supported in part by WPI program (MEXT, Japan), by JSPS Program for Advancing Strategic International Networks to Accelerate the Circulation of Talented Researchers, by JSPS Grant-in-Aid for Scientific Research No.\ 15K17634, by JSPS-NRF research fund, and by Adler Family Fund.

\appendix

\section{Jacobi theta functions}\label{app.Jacobi}

In terms of the following four functions
\begin{align}
\begin{split}
&\ovth_1(x\,|\,\nomec):= 2 \, \EXP^{\ii\pi \frac{\nomec}{4}} \sin(x) \prod_{n=1}^{\infty} (1-\EXP^{2\ii x} \EXP^{\pi \ii \nomec (2n)}) (1-\EXP^{-2\ii x} \EXP^{\pi \ii \nomec (2n)})\,, 
\\[-0.1cm]
&\ovth_2(x\,|\,\nomec):= 2 \, \EXP^{\ii\pi \frac{\nomec}{4}} \cos(x) \prod_{n=1}^{\infty} (1+\EXP^{2\ii x} \EXP^{\pi \ii \nomec (2n)}) (1+\EXP^{-2\ii x} \EXP^{\pi \ii \nomec (2n)})\,, 
\\[-0.1cm]
&\ovth_3(x\,|\,\nomec):= \prod_{n=1}^{\infty} (1+\EXP^{2\ii x} \EXP^{\pi \ii \nomec (2n-1)}) (1+\EXP^{-2\ii x} \EXP^{\pi \ii \nomec (2n-1)})\,, 
\\[-0.1cm]
&\ovth_4(x\,|\,\nomec):=  \prod_{n=1}^{\infty} (1-\EXP^{2\ii x} \EXP^{\pi \ii \nomec (2n-1)}) (1-\EXP^{-2\ii x} \EXP^{\pi \ii \nomec (2n-1)})\,, \label{theta_def}
\end{split}
\end{align}
the Jacobi theta functions $\vartheta_i(x\,|\,\nomec)$, $i=1,\ldots,4$, are defined as
\begin{align}
&\vartheta_i(x\,|\,\nomec):= G(\nomec) \, \ovth_i(x\,|\,\nomec) \quad (i=1,\ldots, 4)\,,
\label{jtheta_def}
\end{align}
where
\begin{align}
\label{qpochdef}
G(\nomec):=\prod_{n=1}^{\infty} (1-\EXP^{2\pi \ii \nomec n}) \,.
\end{align}
In this paper the expressions \eqref{theta_def} are used more often than the expressions \eqref{jtheta_def}.

The four different theta functions in \eqref{theta_def} are related to each other by simple shifts of $x$
\begin{align}
\bs
&\ovth_1(x\,|\,\nomec)=-\ii\EXP^{\ii x+\pi\ii\frac{\nomec}{4}}\,\ovth_4 \left(x+\frac{\pi\nomec}{2}\,|\,\nomec\right), \\
&\ovth_2(x\,|\,\nomec)=\ovth_1\left(x+\frac{\pi}{2}\,|\,\nomec\right),\\
&\ovth_3(z\,|\,\nomec)=\ovth_4\left(x+\frac{\pi}{2}\,|\,\nomec\right).
\es
\end{align}

For this paper we need to use modular transformation properties of the Jacobi theta functions.
For the $T$-transformation, we have
\begin{align}
\bs
\vartheta_1(z \,\big|\, \nomec+1) &= \exp\left(\frac{\pi \ii}{4}\right) \vartheta_1(z \,\big|\, \nomec)\,, \\
\vartheta_2(z \,\big|\, \nomec+1) &=  \exp\left(\frac{\pi \ii}{4}\right) \vartheta_2(z \,\big|\, \nomec)\,, \\
\vartheta_3(z \,\big|\, \nomec+1) &= \vartheta_4(z \,\big|\, \nomec)\,, \\
\vartheta_4(z \,\big|\, \nomec+1) &= \vartheta_3(z \,\big|\, \nomec)\,,
\es
\end{align}
and for the $S$-transformation
\begin{align}
\bs
\vartheta_1\left(z\,\Big|\,-\frac{1}{\nomec}\right)& =
-\ii \sqrt{\frac{\nomec}{\ii}} \exp\left(\frac{\ii}{\pi} \nomec z^2 \right)
\vartheta_1(z \nomec \,\big|\, \nomec)\,, \label{theta1_strans} \\
\vartheta_2\left(z\,\Big|\,-\frac{1}{\nomec}\right)& =
\sqrt{\frac{\nomec}{\ii}} \exp\left(\frac{\ii}{\pi} \nomec z^2 \right)
\vartheta_4(z \nomec \,\big|\, \nomec)\,, \\
\vartheta_3\left(z\,\Big|\,-\frac{1}{\nomec}\right)& =
\sqrt{\frac{\nomec}{\ii}} \exp\left(\frac{\ii}{\pi} \nomec z^2 \right)
\vartheta_3(z \nomec \,\big|\, \nomec)\,, \\
\vartheta_4\left(z\,\Big|\,-\frac{1}{\nomec}\right)& =
\sqrt{\frac{\nomec}{\ii}} \exp\left(\frac{\ii}{\pi} \nomec z^2 \right)
\vartheta_2(z \nomec \,\big|\, \nomec) \,.
\es
\end{align}
By combining these we derive the transformation properties under the
$STS$-transformation:
\begin{align}
\bs
\vartheta_1\left(z\,\Big|\,\frac{\nomec}{1-\nomec}\right)& =
-  \exp\left(\frac{\pi \ii}{4}\right) \sqrt{1-\nomec} \exp\left(\frac{\ii}{\pi} z^2 (\nomec-1) \right)
\vartheta_1\left(z (\nomec-1) \,\big|\, \nomec \right) \,, \\
\vartheta_2\left(z\,\Big|\,\frac{\nomec}{1-\nomec}\right)& =
 \sqrt{1-\nomec} \exp\left(\frac{\ii}{\pi} z^2 (\nomec-1) \right)
\vartheta_3\left(z (\nomec-1) \,\big|\, \nomec \right) \;, \\
\vartheta_3\left(z\,\Big|\,\frac{\nomec}{1-\nomec}\right)& =
 \sqrt{1-\nomec} \exp\left(\frac{\ii}{\pi} z^2 (\nomec-1) \right)
\vartheta_2\left(z (\nomec-1) \,\big|\, \nomec \right) \;, \\
\vartheta_4\left(z\,\Big|\,\frac{\nomec}{1-\nomec}\right)& = \exp\left(\frac{\pi \ii}{4}\right)
 \sqrt{1-\nomec} \exp\left(\frac{\ii}{\pi} z^2 (\nomec-1) \right)
\vartheta_4\left(z (\nomec-1) \,\big|\, \nomec \right) \;. 
\es
\end{align}
Variations of the above equations are also used, such as
\begin{align}
\vartheta_4\left(z\,\Big|\,\frac{\nomec}{N(1-\nomec)}\right)& = \exp\left(\frac{\pi \ii N}{4}\right)
 \sqrt{1-\nomec} \exp\left(\frac{\ii N }{\pi} z^2 (\nomec-1) \right)
\vartheta_4\left(z (\nomec-1) \,\big|\, \frac{\nomec}{N} \right) \;.
\end{align}

\bibliographystyle{nb}
\bibliography{chiral}

\end{document}